\documentclass[prd,onecolumn,aps,superscriptaddress,nofootinbib,showpacs]{revtex4}
\pdfoutput=1
\usepackage{amsmath}
\usepackage{amssymb}
\usepackage{graphicx}
\usepackage{color}
\usepackage{epsfig}
\usepackage{dcolumn}
\usepackage{float}
\usepackage{lipsum}
\usepackage{soul}
\usepackage{bm}
\usepackage{times}
\usepackage{multirow}
\usepackage{hyperref}
\hypersetup{
  colorlinks=true,
  citecolor=blue,
  linkcolor=blue,
  urlcolor=blue}

\usepackage{epsfig}
\usepackage{dcolumn}
\usepackage{morefloats}

\newcommand{\ba}{\begin{array}}
\newcommand{\ea}{\end{array}}
\newcommand{\be}{\begin{equation}}
\newcommand{\ee}{\end{equation}}
\newcommand{\beqa}{\begin{eqnarray}}
\newcommand{\eeqa}{\end{eqnarray}}
\def\321{$SU(3)\times SU(2)\times U(1)$}

\newcommand{\Upmns}{U_{\rm PMNS}}

\hypersetup{
  colorlinks=true,
  citecolor=blue,
  linkcolor=blue,
  urlcolor=blue}
 
\usepackage{epsfig}
\usepackage{dcolumn}
\def\be{\begin{equation}}
\def\ee{\end{equation}}
\def\bea{\begin{eqnarray}}
\def\eea{\end{eqnarray}}
\def\gsim{\ \rlap{\raise 2pt\hbox{$>$}}{\lower 2pt \hbox{$\sim$}}\ }
\def\lsim{\ \rlap{\raise 2pt\hbox{$<$}}{\lower 2pt \hbox{$\sim$}}\ }
\def\dslash{\kern-4pt \not{\hbox{\kern-2pt $\partial$}}}
\def\pslash{\not{\hbox{\kern-2pt p}}}

\newcommand{\dcp}{\delta_{CP}}
\newcommand{\nova}{NO$\nu$A\ }

\begin{document}

\renewcommand{\arraystretch}{2}
\DeclareGraphicsExtensions{.eps,.pdf}


\title{Exploring Partial $\mu$-$\tau$ Reflection Symmetry at DUNE and Hyper-Kamiokande}



\author{Kaustav Chakraborty}
\email[Email Address: ]{kaustav@prl.res.in}
\affiliation{Theoretical Physics Division, Physical Research Laboratory, Ahmedabad - 380009, India}
\affiliation{Discipline of Physics, Indian Institute of Technology, Gandhinagar - 382355, India}

\author{K. N. Deepthi}
\email[Email Address: ]{deepthi@prl.res.in}
\affiliation{Theoretical Physics Division, 
Physical Research Laboratory, Ahmedabad - 380009, India}
  
\author{Srubabati Goswami}
\email[Email Address: ]{sruba@prl.res.in}
\affiliation{Theoretical Physics Division, 
Physical Research Laboratory, Ahmedabad - 380009, India}

\author{Anjan S. Joshipura}
\email[Email Address: ]{anjan@prl.res.in}
\affiliation{Theoretical Physics Division, 
Physical Research Laboratory, Ahmedabad - 380009, India}

\author{Newton Nath}
\email[Email Address: ]{newton@ihep.ac.cn}
\affiliation{Institute of High Energy Physics, Chinese Academy of Sciences, Beijing 100049, China}
\affiliation{School of Physical Sciences, University of Chinese Academy of Sciences, Beijing 100049, China}

%

\begin{abstract}

We study origin, consequences and testability of a hypothesis of 
 `partial $\mu$-$\tau$' reflection symmetry. This symmetry predicts $ |U_{\mu i}|=|U_{\tau i}|~(i=1,2,3) $ for a  single column of the leptonic mixing matrix $U$. Depending on whether 
this symmetry holds for the first or second column of $U$ different 
correlations between $\theta_{23}$ and $ \delta_{CP} $  can be obtained. 
This symmetry can be obtained using discrete flavour symmetries. 
In particular,  all the subgroups of SU(3) with 3-dimensional irreducible 
representation which are classified as class C or D can lead to partial $\mu$-$\tau$ reflection symmetry.
We show how the predictions of this symmetry  compare with the allowed area in the $\sin^2\theta_{23} - \dcp$ plane as obtained from the global analysis of neutrino oscillation data. 
Furthermore, we study the possibility of testing these symmetries at the proposed DUNE and Hyper-Kamiokande (HK) experiments (T2HK, T2HKK),
by incorporating the correlations between $\theta_{23}$ and $ \delta_{CP}$ predicted by the symmetries. 
We find that when simulated data of DUNE and HK 
is fitted with the symmetry predictions, the $\theta_{23}-\dcp$  parameter space gets largely restricted near the CP conserving values of $ \delta_{CP} $. 
Finally, we illustrate the capability of these experiments to distinguish between the 
two cases leading to partial $\mu-\tau$ symmetry namely $|U_{\mu1}| = |U_{\tau 1}|$ and $|U_{\mu 2}| = |U_{\tau 2}|$. 
\end{abstract}
\maketitle

\section{Introduction}
Considerable theoretical and experimental efforts are being devoted towards
predicting and determining the unknowns of the leptonic sectors namely CP
violating phase, octant of the atmospheric mixing angle $\theta_{23}$ (i.e. $\theta_{23}< 45^\circ$,  named as lower octant (LO) or $\theta_{23}> 45^\circ$  named as upper octant (HO)) and
neutrino mass hierarchy (i.e. the sign of $ \Delta m^{2}_{31} $, $ \Delta m^{2}_{31} > 0$  known as normal hierarchy (NH) and $ \Delta m^{2}_{31} < 0$  known as inverted hierarchy (IH)). Symmetry based approaches have been quite
successful in  predicting the interrelations among these quantities 
and the structure of the leptonic mixing matrix as discussed in Refs.
\cite{Altarelli:2010gt, Altarelli:2012ss,Smirnov:2011jv,Ishimori:2010au, King:2013eh} and the references therein. 
General approaches along this line assume some individual residual symmetries of the
leptonic mass matrices which could arise from the breaking of some bigger
symmetry of the leptonic interactions. One such symmetry, called
$\mu$-$\tau$ reflection symmetry, originally discussed by Harrison and Scott in Ref.~\cite{Harrison:2002et} leads to very successful predictions of
mixing angles which are close to the present experimental knowledge.
This symmetry may be stated as equality of moduli of the leptonic mixing
matrix $U$:
\be \label{mtr}
|U_{\mu i}|=|U_{\tau i}|~, \ee
for all the columns $i=1,2,3$. 
Both the origin and consequences of this relation have  been discussed 
in \cite{Grimus:2003yn,Ferreira:2012ri, Grimus:2012hu,
 Mohapatra:2012tb, Ma:2015gka, Joshipura:2015dsa,
Joshipura:2015zla, Joshipura:2016hvn, Nishi:2016wki,
Zhao:2017yvw,Liu:2017frs,Xing:2017mkx,Xing:2017cwb,Joshipura:2018rit,Nath:2018hjx,Zhao:2018vxy}.


Using the standard PDG \cite{Patrignani:2016xqp} parameterization of the matrix $U$
\begin{align}\label{u}
U & =U(\theta_{23}) U(\theta_{13},\delta_{CP})U(\theta_{12}) \nonumber \\
& = \begin{bmatrix}
c_{12} c_{13} & s_{12} c_{13} & s_{13} e^{-i\delta_{CP}} \\
- s_{12} c_{23} - c_{12} s_{23} s_{13} e^{i \delta_{CP}} & c_{12} c_{23} - s_{12} s_{23} s_{13} e^{i \delta_{CP}} & s_{23} c_{13}\\
s_{12} s_{23} - c_{12} c_{23} s_{13} e^{i \delta_{CP}} & - c_{12} s_{23} - s_{12} c_{23} s_{13} e^{i \delta_{CP}} & c_{23} c_{13}
\end{bmatrix}
\end{align}
one finds two well-known predictions
\be \label{mtpredict}
\theta_{23}=\frac{\pi}{4}~,~~ s_{13} \cos\delta_{CP}=0~.\ee
Eq.~(\ref{mtpredict}) suggests maximal $\theta_{23}$, which is allowed within 1$\sigma$
 by the global fits to neutrino observables \cite{deSalas:2017kay,Capozzi:2016rtj,Esteban:2016qun}. Additionally, it allows a nonzero $\theta_{13}$ unlike the
simple $\mu$-$\tau$ symmetry which predicts vanishing $\theta_{13}$
\cite{Fukuyama:1997ky,Ma:2001mr,Lam:2001fb,Balaji:2001ex,Grimus:2004cc,
Joshipura:2005vy}, see recent review \cite{Xing:2015fdg} and references
therein. Here, for $\theta_{13}\not=0$, one gets $\delta_{CP}=\pm\frac{\pi}{2}$
 using eq.~(\ref{mtpredict}). Both these
predictions are
in accord with the global fit of all neutrino data. 
However  a sizeable range is still allowed at 3$\sigma$. 
Note that the best fit value of $\theta_{23}$ in the global fit deviates from the
maximal value for either mass hierarchy. Such deviations can be regarded
as a signal for the departure from the $\mu$-$\tau$ reflection  symmetry. 
A theoretically well-motivated possibility is to assume a `partial
$\mu$-$\tau$' reflection symmetry \cite{Xing:2014zka} and assume that
eq.~(\ref{mtr}) holds only
for a single column\footnote{If it holds for any two columns then by
unitarity, it holds for the third as well.} of $U$. Assuming that it holds
for the third column, one gets maximal $\theta_{23}$ and $\delta_{CP}$ remains
unrestricted.
These correlations are found from eq.~(\ref{u}) in respective cases $i=1$ and $i=2$ to be 

\begin{eqnarray}
\cos \delta_{CP}  & = \frac{(c^2_{23} - s^2_{23})(c^2_{12} s^2_{13} - s^2_{12}
)}{4 c_{12}s_{12}c_{23}s_{23}s_{13}}, ~~~(|U_{\mu 1}|=|U_{\tau 1}|
)~~~ {\rm , } ~~~C_1\label{eq:model_1}~,\\
\cos \delta_{CP}  & = \frac{(c^2_{23} - s^2_{23})(c^2_{12} - s^2_{12}
s^2_{13})}{4 c_{12}s_{12}c_{23}s_{23}s_{13}}, ~~~(|U_{\mu 2}|=|U_{\tau 2}| ) ~~~
{\rm , } ~~~ C_2 \label{eq:model_2}~. 
\end{eqnarray}

These equations correlate the sign of $\cos\delta_{CP}$ to the octant of
$\theta_{23}$. $\theta_{23}$ in the first (second) octant leads to a
negative (positive) value of $\cos\delta_{CP}$ in case of
eq.~(\ref{eq:model_1}). 
It predicts exactly opposite behaviour for eq.~(\ref{eq:model_2}). 
The exact quadrant of $\delta$ is still not fixed by these equations but it
can also be determined from symmetry considerations\cite{Joshipura:2018rit}.
These correlations were also obtained in \cite{Ge:2011ih,Ge:2011qn} in the context of
$Z_2$ and $\overline{Z_2}$ symmetries 
\footnote{See, the review article \cite{Petcov:2017ggy} for references on other similar sum rules and their testability.}. 
Henceforth, we refer to these correlations as $C_1$ and $C_2$ respectively.
The above equations also indirectly lead to information on the
neutrino mass hierarchy since the best fit values of $\theta_{23}$ lie in the first
(second) octant in case of the normal (inverted) hierarchy according to
the latest global fits reported in \cite{Capozzi:2016rtj,Esteban:2016qun,deSalas:2017kay}. Thus precise verification of the above equations is of considerable importance and the long baseline
experiments can provide a way for such study. 
Similar study has been  performed in the context of the \nova and T2K experiments
in \cite{Toorop:2011jn,Hanlon:2013ska,Hanlon:2014bga}. 

In this paper, we consider the testability of these relations at the 
forthcoming long baseline experiments 
Deep Under-ground Neutrino Experiment (DUNE) and Hyper-Kamiokande (HK). 
These potential high-statistics experiments will overcome the parameter
degeneracies faced by the current experiments and lead us in to an era 
of precision measurements of the oscillation parameters \cite{Acciarri:2015uup,raj_lbne1,suprabhlbnelbno,Barger:2014dfa,
Bora:2014zwa,Ghosh:2014rna,Dutta:2014yra,Deepthi:2014iya,Nath:2015kjg,Srivastava:2018ser}.  
Because of this, these experiments are ideal to test the parameter
correlations  like the ones given in the eqs.~(\ref{eq:model_1}, \ref{eq:model_2}).  
In the following, we obtain the allowed parameter range in the $\dcp$ -
$\sin^2\theta_{23}$ plane by fitting the symmetry relations embodied 
in the eqs.~(\ref{eq:model_1}, \ref{eq:model_2}) to the simulated DUNE and HK data. 
We also discuss whether the correlations  $C_1$(eq.~(\ref{eq:model_1})) and $C_2$(eq.~(\ref{eq:model_2})) can be distinguished at DUNE and HK. 
Recent studies on testing various models from future experiments can be found
for instance in \cite{Pasquini:2018udd,Srivastava:2017sno,Agarwalla:2017wct,Chatterjee:2017ilf,Petcov:2018snn,Ballett:2016yod,Ballett:2015wia,Ballett:2014dua,Penedo:2017knr}.

We begin by first discussing the origin of partial $\mu$-$\tau$ reflection
symmetry, after which in Section~\ref{sec:part_symm} we elaborate on the robustness of the resulting predictions in a large
class of models based on flavour symmetry.
We give a brief overview of the experiments and simulation details in Section~\ref{sec:expt_speci}. In  Section~\ref{sec:result}, we perform 
a phenomenological analysis of the testability of the above symmetries 
in DUNE and HK. We use the extra correlations predicted by the symmetry 
in fitting the simulated data of these experiments and obtain the allowed areas in the 
$\dcp-\sin^2\theta_{23}$ plane.  
In subsection \ref{sec:result_model_diff}, we discuss the possibility of 
differentiating between the two symmetries -- $C_1$ and
$C_2$.
We draw our conclusions in Section~\ref{sec:conclusion}.

\section{Partial $\mu$-$\tau$ Reflection Symmetry and Discrete Flavour Symmetries}\label{sec:part_symm}
We briefly review here the  general approach based on flavour symmetry to
emphasize that partial $\mu$-$\tau$ reflection symmetry is  a generic
prediction of almost all such schemes barring few exceptions. Basic
approaches assume  groups $G_\nu$ and $G_l$ as the residual symmetries of
the neutrino mass matrix $M_\nu$ and the charged lepton mass matrix
$M_lM_l^\dagger$ respectively. Both these groups are assumed to arise from
the breaking of some unitary discrete group $G_f$. The $\Upmns$ matrix $U$
gets fixed upto
the neutrino Majorana phases if it is further assumed that $G_\nu=Z_2\times
Z_2$ and $G_l=Z_n,n\geq 3$. In addition, if we demand that all the predicted
mixing angles are non-zero, then the following unique form is predicted for
almost all the discrete groups $G_f$ \cite{King:2014rwa,Joshipura:2016quv}
\be \label{ugen}
U\equiv U_{\rm gen}(\theta_n)=\frac{1}{\sqrt{3}}\left(
\ba{ccc}
\sqrt{2} \cos\theta_n&1&\sqrt{2}\sin\theta_n\\
\sqrt{2}
\cos(\theta_n-\frac{2\pi}{3})&1&\sqrt{2}\sin(\theta_n-\frac{2\pi}{3})\\
\cos(\theta_n-\frac{4\pi}{3})&1&\sqrt{2}\sin(\theta_n-\frac{4\pi}{3})\\
\ea\right)~,\ee
where $\theta_n\equiv \frac{\pi a }{n}$ is a discrete angle with
$a=0,1,2....\frac{n}{2}$. We have not shown here the unphysical phases which can be absorbed in defining charged lepton fields and unpredicted Majorana phases.
All the discrete subgroups of $SU(3)$ with three dimensional irreducible
representation are classified as class $C$ or $D$ and five
exceptional groups \cite{Grimus:2013apa}. Eq.~(\ref{ugen}) follows in all the type $D$ groups
taken as $G_f$. Type $C$ groups lead instead to democratic mixing which shows full $\mu$-$\tau$ reflection symmetry but predict large reactor angle. Eq.~(\ref{ugen}) arises even if $G_f$ is chosen as a
discrete subgroup of $U(3)$ having the same textures as class $D$
groups \cite{Joshipura:2016quv}. 

Eq.~(\ref{ugen}) displays partial $\mu$-$\tau$  reflection symmetry for the
second column for all the values of $\theta_n\not=0,\frac{\pi}{2}$. In the
latter case, one gets total $\mu$-$\tau$  reflection symmetry but at the
same time
one of the mixing angles is predicted to be zero and one would need to break
the
assumed residual symmetries to get the correct mixing angles. More importantly, eq.~(\ref{ugen}) being essentially a real matrix also predicts trivial
Dirac CP phase $\delta_{CP}=0$ or $\pi.$ 
{Eq.~(\ref{eq:model_2}) in this case implies  a correlation among angles. 
}  
Non-zero CP phase and 
partial $\mu$-$\tau$  symmetry in other columns can arise in an alternative but less
predictive
approach in which the residual symmetry of the neutrino mass matrix is taken
as $Z_2$ instead of $Z_2\times Z_2$. In this case, one can obtain the following  mixing matrix $U$ with a proper choice of residual symmetries
\be \label{z2miz}
U=U_{\rm gen}(0) U_{ij}~,\ee
where $U_{ij}$ denotes a unitary rotation either in the  $ij^{th}$ plane
corresponding to partial symmetry in the $k^{th}$($i\not=j\not=k$) column . Examples of 
the required residual symmetries are discussed in \cite{Altarelli:2010gt, Altarelli:2012ss,Smirnov:2011jv,Ishimori:2010au, King:2013eh} and  minimal example of this occurs with $G_f=S_4.$ 

The partial $\mu$-$\tau$ symmetries obtained this way also lead to
additional restrictions 
\be \label{addn_predictn1}
c_{12}^2c_{13}^2=\frac{2}{3} \ee
and
\be \label{addn_predictn2}
s_{12}^2c_{13}^2=\frac{1}{3}~,
\ee
where eq.~(\ref{addn_predictn1}) and eq.~(\ref{addn_predictn2}) 
follow from the partial symmetries of the
first and second columns respectively.
These predictions arise here from the requirement that $G_\nu$
and $G_l$ are embedded in DSG of $SU(3)$ and need not arise in a more
general approach. It is then  possible to obtain  specific symmetries
\cite{Ge:2011ih,Ge:2011qn} in which solar angle is a function of a
continuous parameter.

\noindent \begin{figure}[!h]
\begin{center}
 \begin{tabular}{lr}
\hspace{0.8cm}
\includegraphics[height=5.5cm,width=6.5cm]{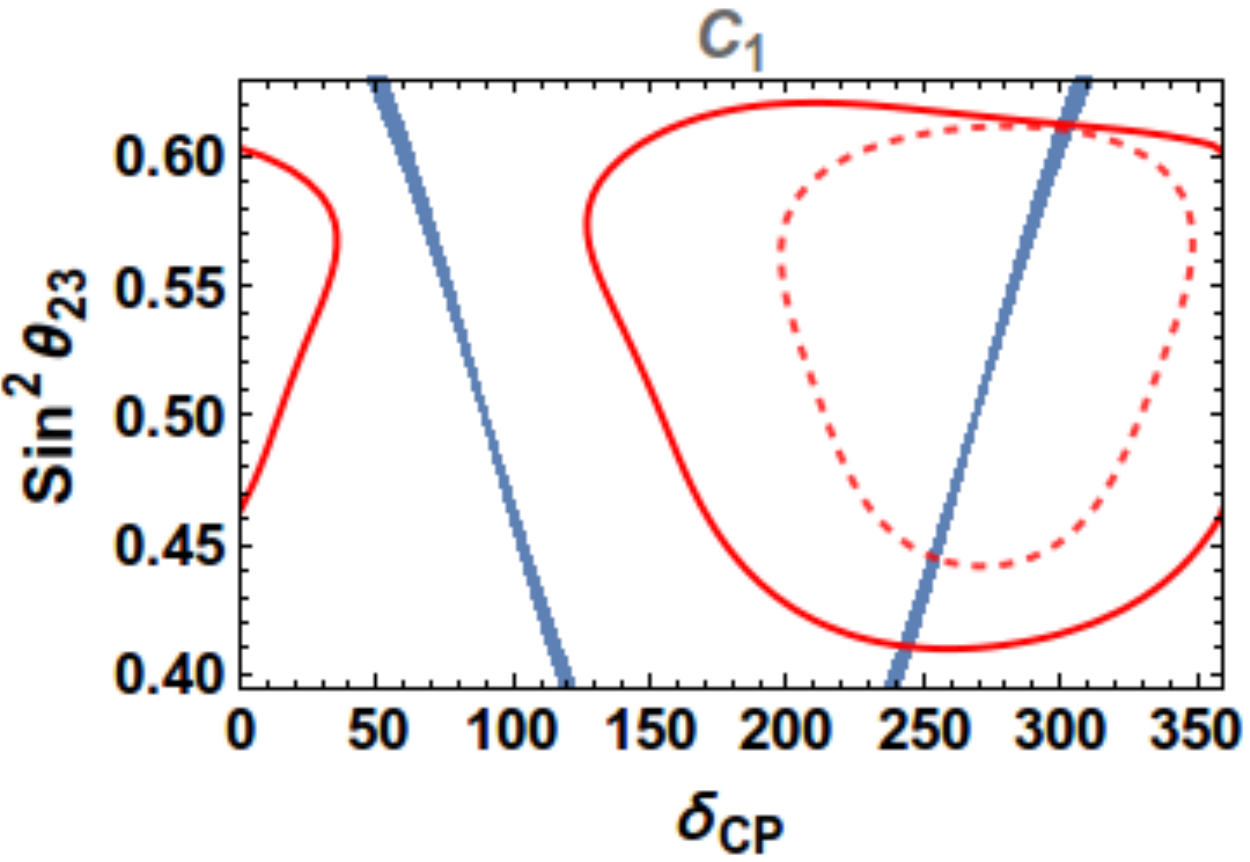}
\hspace{0.40in}
\includegraphics[height=5.5cm,width=6.5cm]{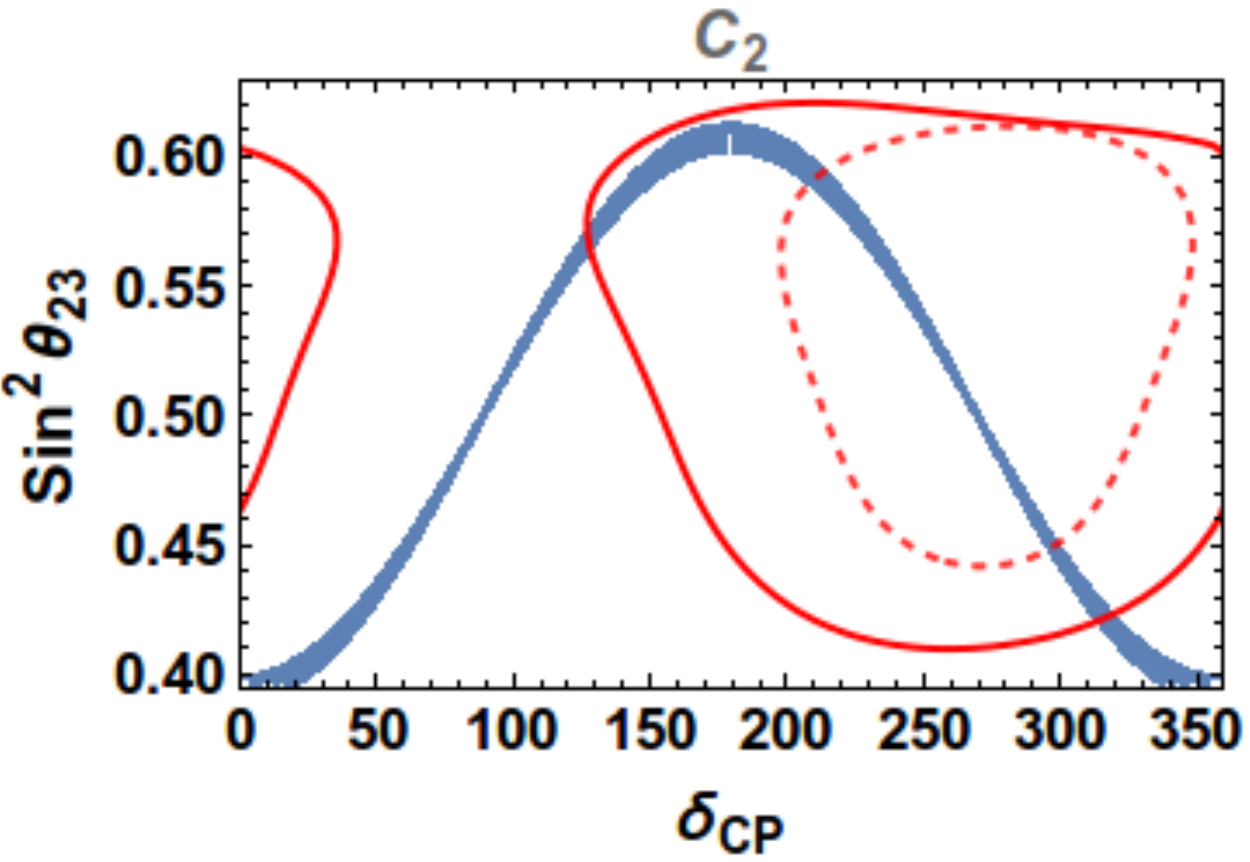}
 \end{tabular}
 \end{center}
\vspace{-6ex}        
\caption{\footnotesize  The thick blue lines show the correlation plots
in the $\sin^2\theta_{23} $  -  $\dcp$  plane as predicted by the symmetry relations. The left (right) panel correspond to Eq.~(\ref{eq:model_1}) (Eq.~(\ref{eq:model_2})). The solid(dashed) red curves represent the 3$\sigma$ allowed parameter space as obtained by the global analysis 
of data by the Nu-fit  collaboration \cite{Esteban:2016qun,nufit16} considering hierarchy to be NH(IH) respectively.}
\label{fig:th23_cp_corr}
\end{figure}

Fig.~\ref{fig:th23_cp_corr}, shows the correlation plots (thick blue lines) between 
$\sin^2\theta_{23}$ and $\dcp$ as given by eqs.~(\ref{eq:model_1}, \ref{eq:model_2}). 
Here, the red solid(dashed) contours represent the $3\sigma$ allowed region for NH(IH) as obtained from
the global-fit data by the Nu-fit collaboration \cite{nufit16}.
Eqs.~(\ref{eq:model_1}, \ref{eq:model_2}) give two values of CP phase (namely, $\dcp$ and $360^\circ - \dcp$) for each value of $\theta_{23}$ except for $  \dcp \equiv 180^\circ$. 
The width of the blue lines is due to the uncertainty of the angles 
$\theta_{12}$ and $\theta_{13}$ subject to the conditions given in
eq.~(\ref{addn_predictn1}) and eq.~(\ref{addn_predictn2}) corresponding to 
eq.~(\ref{eq:model_1}) and eq.~(\ref{eq:model_2}) respectively. 
It is seen that the correlation between $\sin^2\theta_{23}$ and $\dcp$ 
is opposite in the class of symmetries  that give eq.~(\ref{eq:model_1}) vis-a-vis 
those that give eq.~(\ref{eq:model_2}). 
The parameters, $\sin^2\theta_{23}$ and $\dcp$ are correlated between $0^\circ - 180^\circ$ 
and anti-correlated between $180^\circ - 360^\circ$ for 
eq.~(\ref{eq:model_1}). The opposite is true for 
eq.~(\ref{eq:model_2}). We also notice here that eq.~(\ref{eq:model_2}) rules out regions around CP conserving (i.e. $0^\circ,180^\circ, 360^\circ$) values. 
Additionally we observe that, at $3\sigma$ some of the allowed regions of $\sin^2\theta_{23} $ and $ \dcp$ as predicted by the symmetries are disfavoured by the current global-fit data. From the global-fit data we observe that the region $39^\circ < \dcp < 125^\circ $ is completely ruled out at $3\sigma$ for NH and the region $ \dcp < 195^\circ $ for IH.
The symmetry predictions can further constrain the values of $\dcp$
presently allowed
by the global data. 

In the next section, we study how far the allowed areas 
in $\dcp - \sin^2 \theta_{23}$ plane can be restricted 
if the simulated experimental data confronts the symmetry predictions.

\section{Specifications of the Experiments}\label{sec:expt_speci}\label{sec:expt_spe}
In this paper, we have simulated all the experiments
using the \texttt{GLoBES} package \cite{globes1,globes2} along with the required auxiliary files~\cite{messier_xsec,paschos_xsec}. We have considered
the experimental set-up and the detector performance of DUNE and HK in accordance with
 ref. \cite{Alion:2016uaj} and ref. \cite{Abe:2016ero} respectively.

\begin{itemize}
\item \textbf{Deep Underground Neutrino Experiment (DUNE) :}
DUNE is a Fermilab based next generation long baseline superbeam experiment. 
This experiment will utilize upcoming leading edge facility --
Long Baseline Neutrino Facility (LBNF), which will provide a
high intensity neutrino beam and the infrastructure required for the DUNE
experiment.
In this experiment, the muon-neutrino beam from  Fermilab will travel a baseline
of 1300 km before it gets detected at the far detector situated at the 
``Sanford Underground Research Facility (SURF)" in Lead, South Dakota. 
The proposed far detector for DUNE is a LArTPC (liquid argon time-projection chamber) detector with the volume of 40 kT. 
The beam power will be initially 1.2 MW and later will be increased to 
2.3 MW \cite{lbne}. In our simulation we consider the 
neutrino flux \cite{cherdack} 
corresponding to 1.2 MW beam power which gives  $ 1\times 10^{21} $ protons on target (POT) per year. This corresponds to a proton energy of 120 GeV. We also consider a total run time of 
(5$\nu$+5$\bar{\nu}$) years as proposed by the experiment.

\item \textbf{Hyper-Kamiokande Experiment :} 
Hyper-Kamiokande experiment \cite{Abe:2016ero} is a Japan based long baseline experiment 
which will use J-PARC (Japan Proton Accelerator Research Complex) neutrino beam facility. 
The primary goal of HK experiment is to determine CP violation. However, it is also capable of 
observing nucleon decay, atmospheric neutrinos and neutrinos of astronomical origin.
Recently the collaboration has proposed  
two alternatives for the location of the far detector. The first
one is T2HK (Tokai-to-Hyper-Kamiokande) which plans on constructing two water-cherenkov detectors
(cylindrical tanks) of fiducial volume 187 kt at 295 km in Kamioka. Alternatively, T2HKK proposes
to have one tank of 187 kt at 295 km in Kamioka and the other 187 kt tank at 1100 km in Korea  \cite{Abe:2016ero}. 
In our simulations we have considered 
the off-axis angle (OAA) for this detector in Korea as $1.5^\circ$, 
proposed run time ratio to be 1:3 in neutrino and antineutrino modes 
(total run time - 10 years) and the 
proton beam power of 1.3 MW giving a total of $27 \times 10^{21}$ protons on target (POT).
\end{itemize}
 \begin{table}[!h]
 \centering
 \begin{tabular}{|c|c|c|}
 \hline
 Osc. param. & True Values &   Test Values \\
 \hline   
$ \sin^{2} \theta_{13} $ & 0.0219 & 0.0197 -- 0.0244 \\
$ \sin^{2}\theta_{12} $ & 0.306 & 0.272 -- 0.346 \\
$ \theta_{23} $ & 39$ ^\circ -  51 ^\circ $  &  39$ ^\circ -  51 ^\circ$ \\
$\Delta m^2_{21} ({\rm eV^{2}}) $ &  7.50 $ \times 10^{-5} $  &  Fixed \\
$\Delta m^2_{31}({\rm eV^{2}}) $ & 2.50 $ \times 10^{-3}  $ & (2.35 -- 2.65) $ \times 10^{-3} $ \\
$ \delta_{CP} $ &  (0-360)$^\circ $  & {\rm{Symmetry predictions}}  \\  
%
  \hline
 \end{tabular}
\caption{\footnotesize Values of Oscillation parameters that are considered in 
 this study unless otherwise mentioned. We vary the true 
values of $\theta_{23}$ in the whole allowed range and marginalization 
for each $\theta_{23}^{\mathrm{true}}$ is done over the full allowed range
of $\theta_{23}$. See text for more details. 
}
 \label{tab:param_values}
 \end{table}
 
%
\section{Phenomenological Analysis}\label{sec:result}
In this section, we perform a phenomenological analysis 
exploring the possibility of probing the correlations
$C_1$ and $C_2$ at DUNE, T2HK and T2HKK. 
This is discussed in terms of correlation plots in $\sin^2\theta_{23}-\dcp$ plane.
We also discuss the possibility of distinguishing between the two models at these experiments.  

We perform a $ \chi^{2} $ test with $\chi^2$ defined as,
\begin{eqnarray}
\chi^2_{{\rm tot}} = \underset{\xi, \omega}{\mathrm{min}} \lbrace \chi^2_{{\rm stat}}(\omega,\xi) + \chi^2_{{\rm pull}}(\xi) + \chi^2_{prior} \rbrace.
 \label{chi-tot}
\end{eqnarray}
$\chi^2_{{\rm stat}}$ is the  statistical $\chi^2$ whereas 
$\chi^2_{pull}$ signifies the systematic uncertainties which 
are included using the method of pulls with $\xi$ denoting the pull variable 
\cite{pulls_gg, pull_lisi,ushier}. Here, $\omega$ represents the  oscillation parameters :
\{$\sin^2\theta_{23}, \sin^2\theta_{12}, \delta_{CP}, \Delta m^2_{21},\Delta m^2_{31}$\}.   
$\chi^2_{prior}$  captures the knowledge of the oscillation parameters 
from other experiments  and is defined as, 
\begin{equation} 
\chi^2_{prior} (p)= \frac{(p_0 - p)^2}{\sigma_0^2} \;,
\end{equation}
$p$ denotes the parameter for which a prior is added and $p_0$ and
$\sigma_0$  correspond to its best fit  value  and $1\sigma$ error, respectively. 
In our analysis we have considered the effect of prior on the parameters 
$\theta_{13}$ and $\theta_{12}$. 
We assume Poisson distribution to calculate the statistical $ \chi^{2}_{stat} $,
\begin{equation}\label{eq:stat_chisq}
 \chi^{2}_{stat} = \sum_i 2(N^{test}_i-N^{true}_i - N^{true}_i \log\dfrac{N^{test}_i}{N^{true}_i}) \;.
\end{equation}
Here, `i' refers to the number of bins and $ N^{test}_i, N^{true}_i $
are the total number of events due to test and true set of oscillation
parameters respectively.  $N^{test}_i$ is defines as follows, including 
the effect of systematics    
\begin{eqnarray}
N^{{\rm (k)test}}_i(\omega, \xi) = \sum\limits_{k = s, b} N^{(k)}_{i}(\omega)[1 + c_{i}^{(k) norm} \xi^{(k) norm} + c_{i}^{(k) tilt} \xi^{(k) tilt} \frac{E_i - \bar{E}}{E_{max} - E_{min}}] \;,
\end{eqnarray}
where $k = s(b)$ represent the  signal(background) events. 
$c_i^{norm}$(${c_i}^{tilt}$)  corresponds to the change in  the number of events due to the  
 pull variable $\xi^{norm}$(${ \xi}^{tilt}$). In the above equation $E_i$ denotes
the mean reconstructed energy of the $i^{th}$ bin with  $E_{min}$ and $E_{max}$ representing  the maximum and minimum energy in the entire energy range and $\bar{E} = ({E_{max} +E_{min}})/{2}$ is the mean energy over this range. The systematic uncertainties (normalization errors) and efficiencies 
corresponding to signals and backgrounds of DUNE and HK
are taken from \cite{Acciarri:2015uup, Abe:2016ero}. 
For DUNE the signal normalization uncertainties on 
$\nu_e$/$\bar{\nu_{e}}$ and $\nu_{\mu}$/$\bar{\nu_{\mu}}$ 
are considered to be 2\% and 5\% respectively. While a range of 
5\% to 20\% background uncertainty along with the correlations 
among their sources have also been included.
On the other hand, for T2HK the signal normalization 
error on $\nu_e$($\bar{\nu_{e}}$) and $\nu_{\mu}$($\bar{\nu_{\mu}}$)
are considered to be 3.2\% (3.9\%) and 3.6\%(3.6\%) respectively. 
In the case of T2HKK, 3.8\% (4.1\%) and 3.8\% (3.8\%)
are taken as the signal normalization 
errors on $\nu_e$($\bar{\nu_{e}}$) and $\nu_{\mu}$($\bar{\nu_{\mu}}$)
respectively.
The background normalization uncertainties range from 3.8\% to 5\%.
${N_i}^{{\rm true}}$ in eq.~(\ref{eq:stat_chisq}) is obtained by adding the 
the simulated signal and background events i.e.  
${N_i}^{{\rm true}} = N_i^s + N_i^b$.

In Table \ref{tab:param_values}, we list the values for the neutrino
oscillation  parameters that we  have used in our numerical simulation.
These values are consistent with the results  obtained from global-fit
of world neutrino data
\cite{Capozzi:2016rtj,Esteban:2016qun,deSalas:2017kay}.

\subsection{Testing the $\sin^2\theta_{23} - \dcp$ correlation predicted by the symmetries at DUNE, T2HK and T2HKK} \label{sec:correlation}

The numerical analysis is performed as follows. 
\begin{itemize} 
\item 
The  data corresponding to each experiment is generated by considering the true values of the oscillation parameters given in table~\ref{tab:param_values}. Note that the true 
values of $\theta_{23}$ and $\dcp$ are spanned over the range $(39-51)^\circ$  
and  ($0^\circ - 360^\circ$), respectively.  
\item 
In the theoretical fit, we calculate the test events by
marginalizing over the parameters $\sin^2\theta_{13}$, $|\Delta m^2_{31}|$,
$\sin^2\theta_{23}$ and $\sin^2\theta_{12}$ in the test plane using the
ranges presented  in table~\ref{tab:param_values}.  
\item The test values of $\dcp$ used are as predicted by the symmetries  
specified in eq.~(\ref{eq:model_1}) for $C_1$ and eq.~(\ref{eq:model_2}) in $C_2$. 
\item In addition we impose the conditions given in eq.~(\ref{addn_predictn1}) for the 
symmetry relation $C_1$ and eq.~(\ref{addn_predictn2}) for
symmetry relation $C_2$ in the test.  Note that given the current range of $\sin^2\theta_{13}$  
these relations restrict the value of $\sin^2\theta_{12}$ to  $ 0.316 < \sin^2\theta_{12} < 0.319$ for $C_1$ and $\sin^2\theta_{12}$ to $0.34 < \sin^2\theta_{12} < 0.342$ for $C_2$.
These values of $\sin^2\theta_{12}$ are within the current $3\sigma$ allowed range. 
Note that these ranges exclude the current best-fit value of $\sin^2\theta_{12}$.  
Precise measurement of $\theta_{12}$ for instance in the reactor neutrino experiment JUNO  \cite{Li:2016ozg}  can provide a stringent test of these scenarios. 

\item We have not added any prior in this analysis. We have checked that the prior on $\theta_{13}$ does not play any role when the constraints represented by  
eq.~(\ref{addn_predictn1}) and eq.~(\ref{addn_predictn2}) are 
applied. In addition, since the best-fit $\theta_{12}$ 
is excluded already by the constraints, the imposition of $\theta_{12}$ 
prior will disfavour the scenarios.  


\item We minimize the $\chi^2$ and plot the regions in the 
$\sin^2\theta_{23}(\mathrm true) - \dcp (\mathrm true)$ plane for which 
$\chi^2 \leq \chi^2_{min} + \Delta \chi^2$ where $\Delta \chi^2$ values 
used  correspond to 1, 2 and 3 $\sigma$.  

\end{itemize} 
 
\begin{figure}[h!]
\begin{center}
\begin{tabular}{lr}
\vspace{0.2in}
\includegraphics[height=5.8cm,width=6cm]{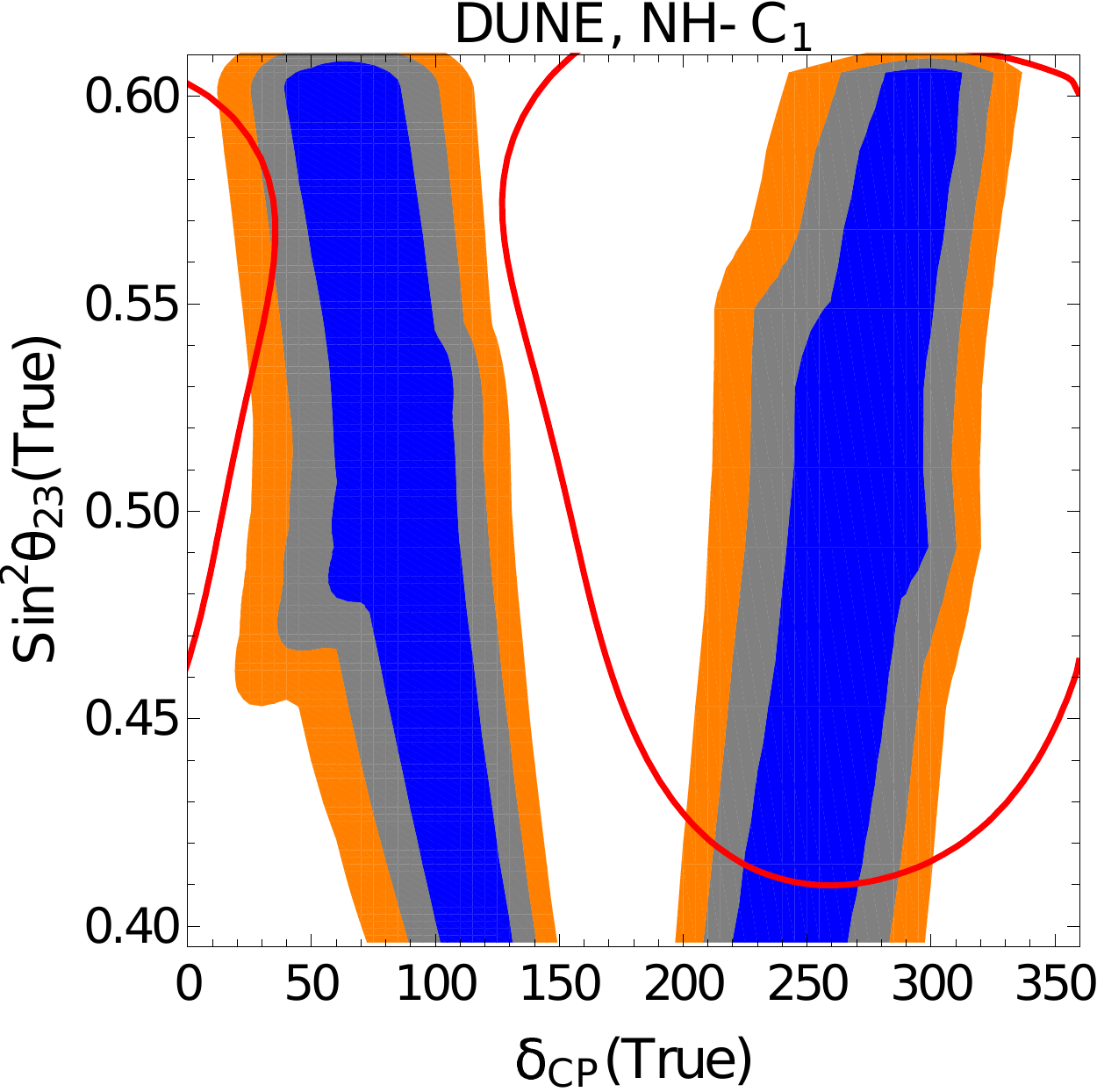}
\hspace{0.5in}
\includegraphics[height=5.8cm,width=6cm]{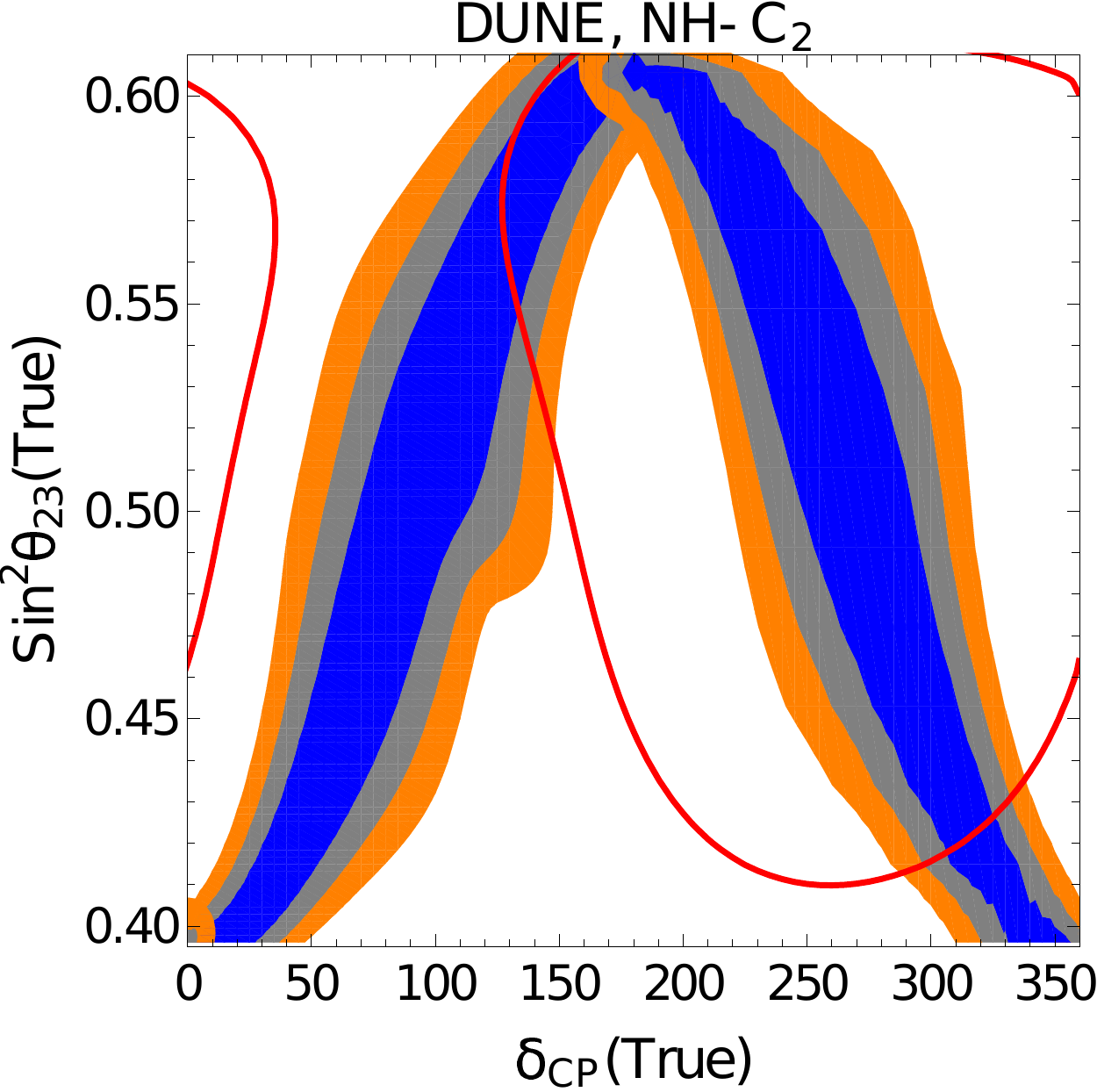}\\
\vspace{0.2in}
\includegraphics[height=5.8cm,width=6cm]{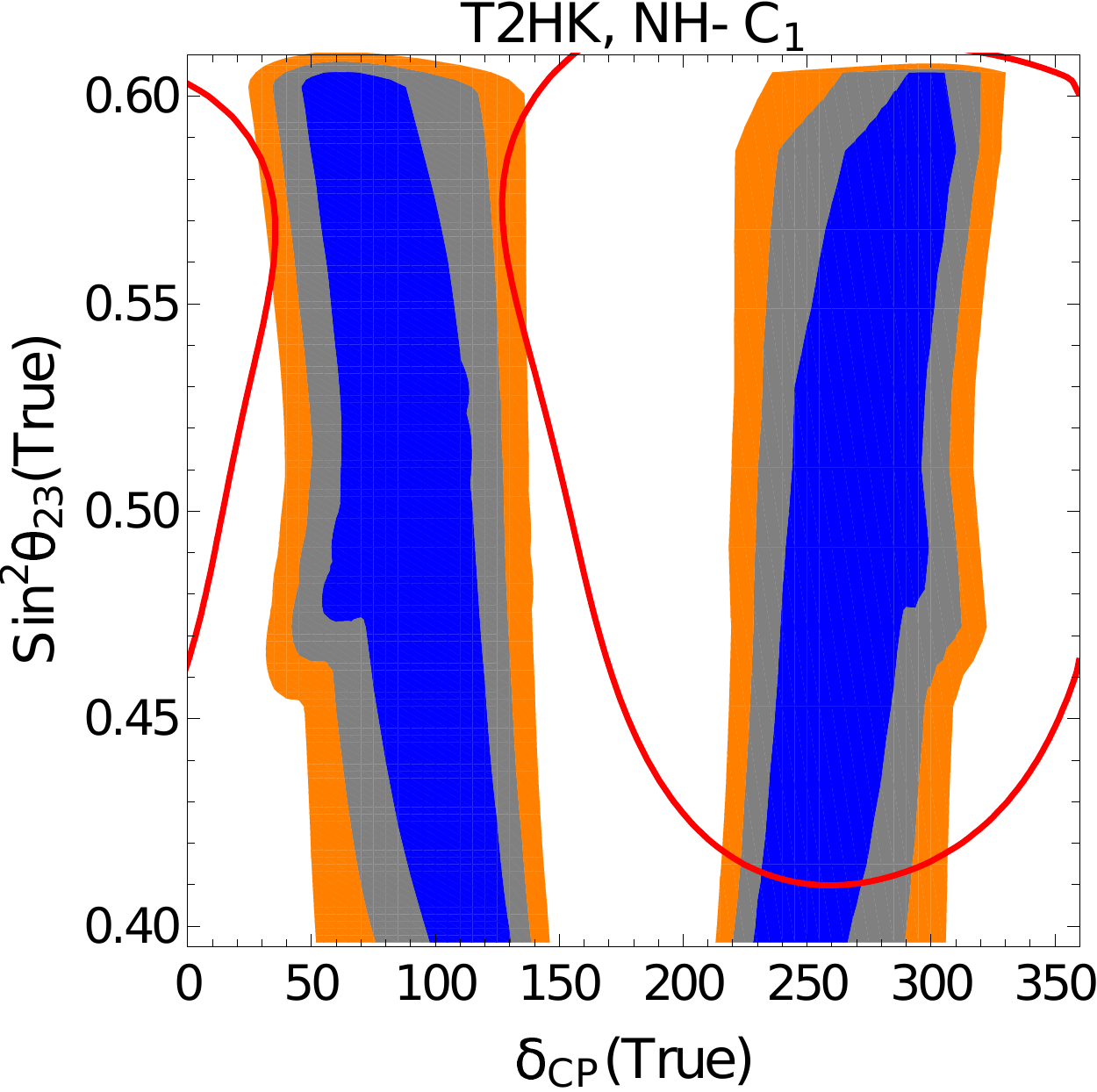}
\hspace{0.5in}
\includegraphics[height=5.8cm,width=6cm]{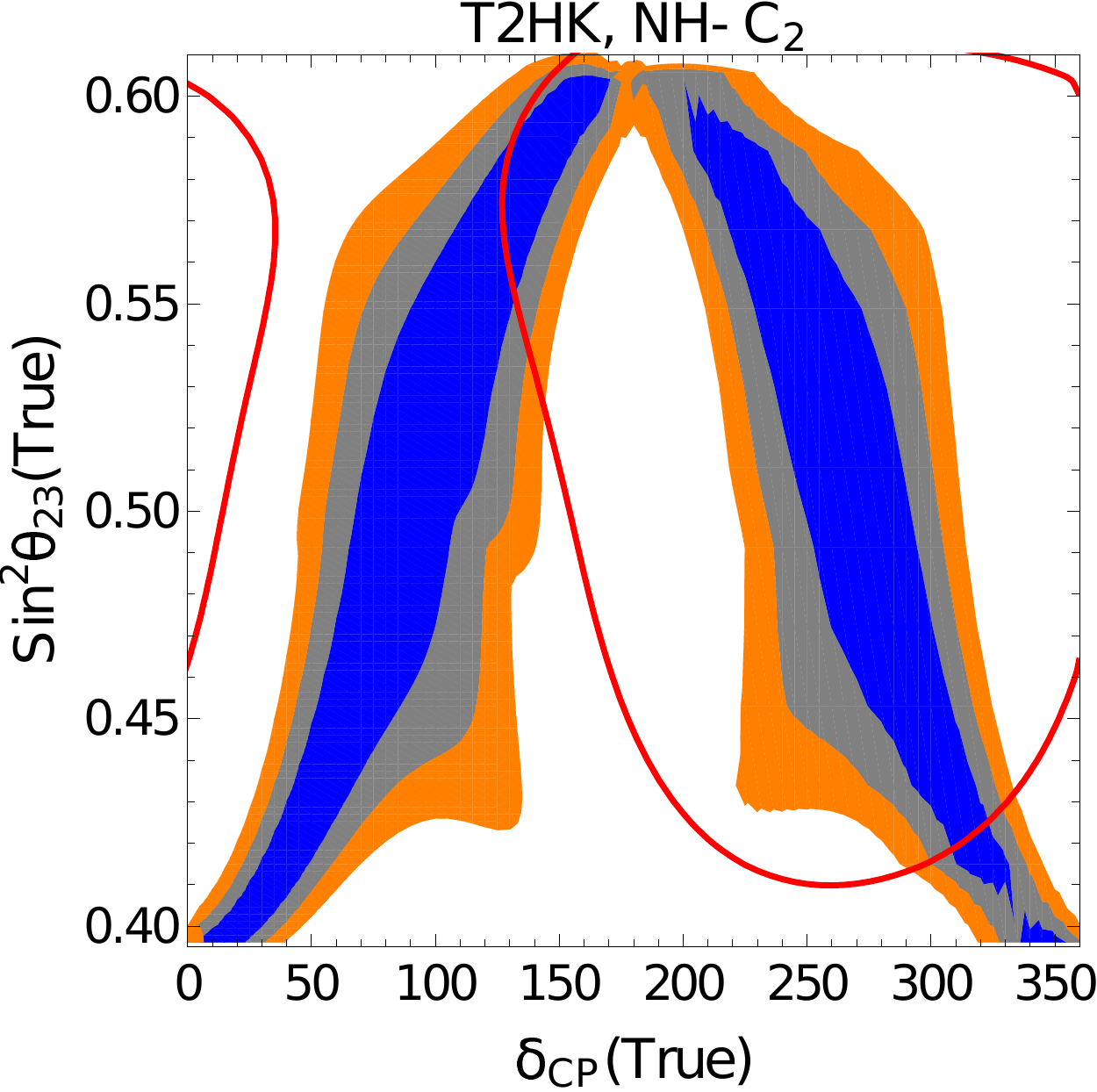} \\
\includegraphics[height=5.8cm,width=6cm]{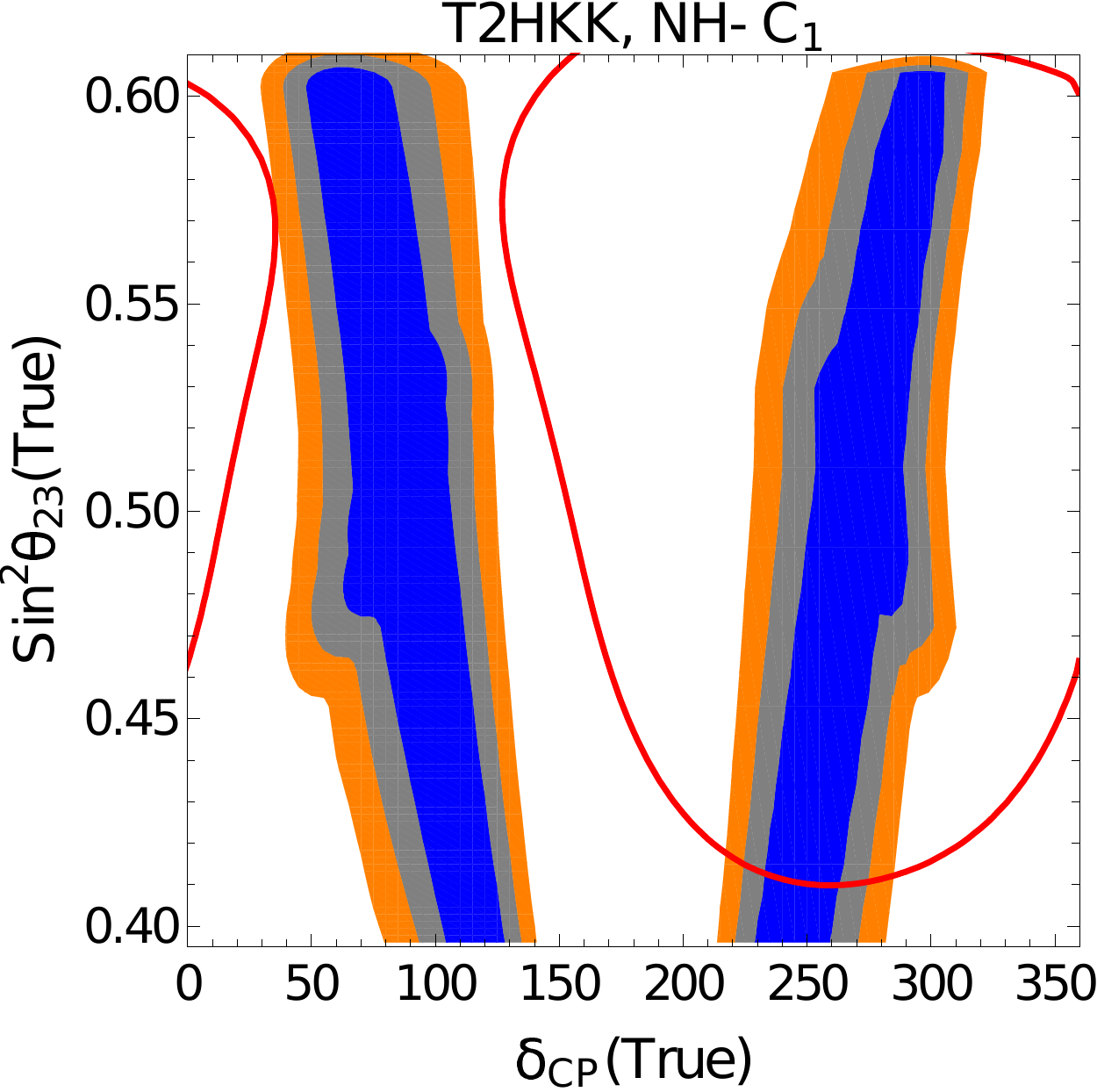}
\hspace{0.5in}
\includegraphics[height=5.8cm,width=6cm]{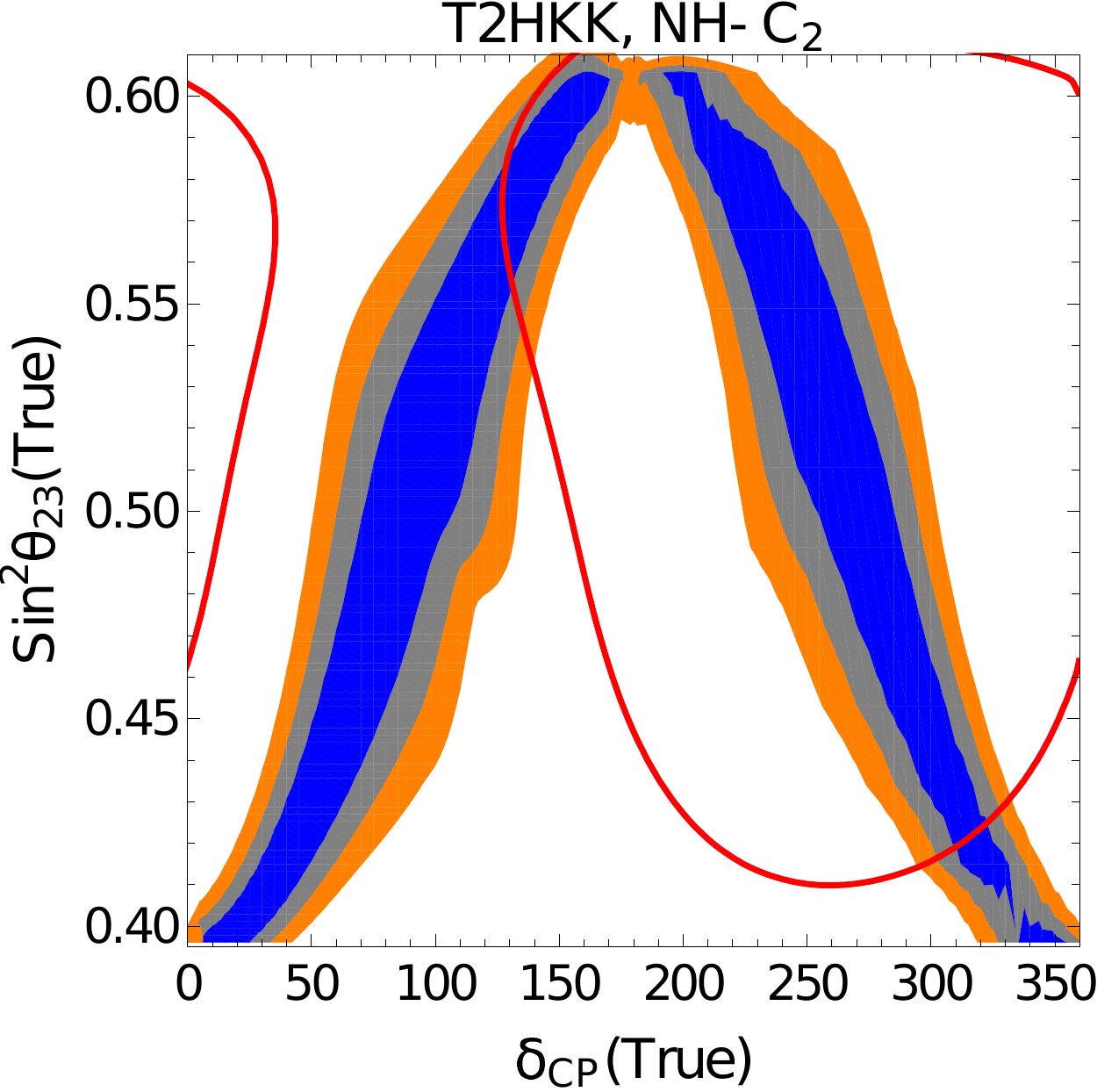} \\
 \end{tabular}
 \end{center}
\vspace{-0.8cm} 
\caption{\footnotesize Contour plots in the true: $\sin^{2} \theta_{23}$(true) - $\dcp$(true) plane for DUNE, T2HK, T2HKK. The left(right) panel represents the prediction from the symmetry relation $C_1$($C_2$) which corresponds to the eq.~(\ref{eq:model_1}) ((eq.~\ref{eq:model_2})). The hierarchy is fixed as NH. The red contour in each panel represents the  $3\sigma$ allowed area from the global analysis of neutrino oscillation data as obtained by the Nu-fit collaboration \cite{Esteban:2016qun,nufit16} for Normal Hierarchy. The blue, gray and yellow shaded contours correspond to 1$ \sigma $, 2$ \sigma $, 3$ \sigma $ respectively.}
\label{fig:th23_cp_bound}       
\end{figure}

\begin{figure}[h!]
\begin{center}
\begin{tabular}{lr}
\vspace{0.2in}
\includegraphics[height=5.8cm,width=6cm]{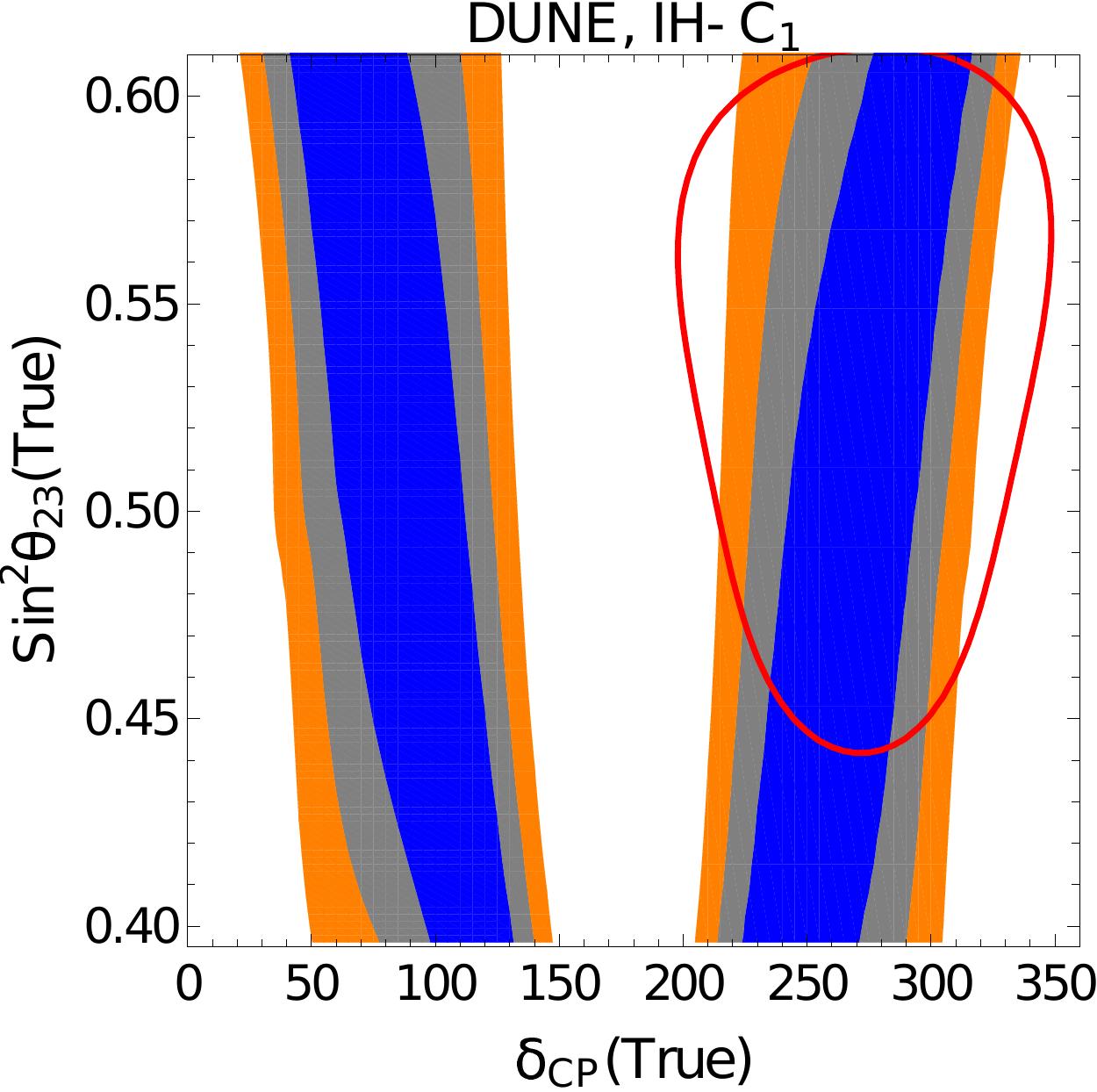}
\hspace{0.5in}
\includegraphics[height=5.8cm,width=6cm]{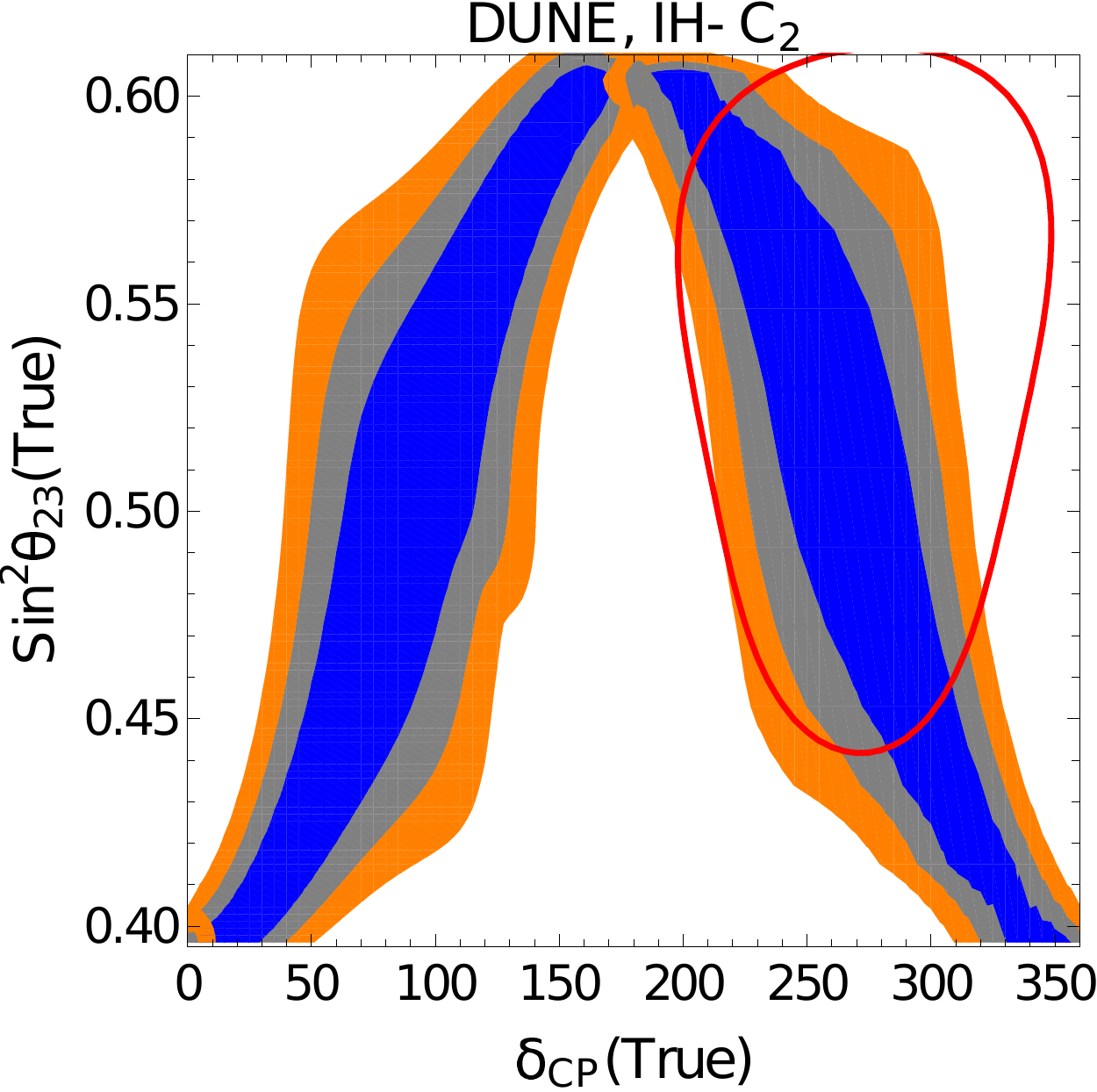}\\
\vspace{0.2in}
\includegraphics[height=5.8cm,width=6cm]{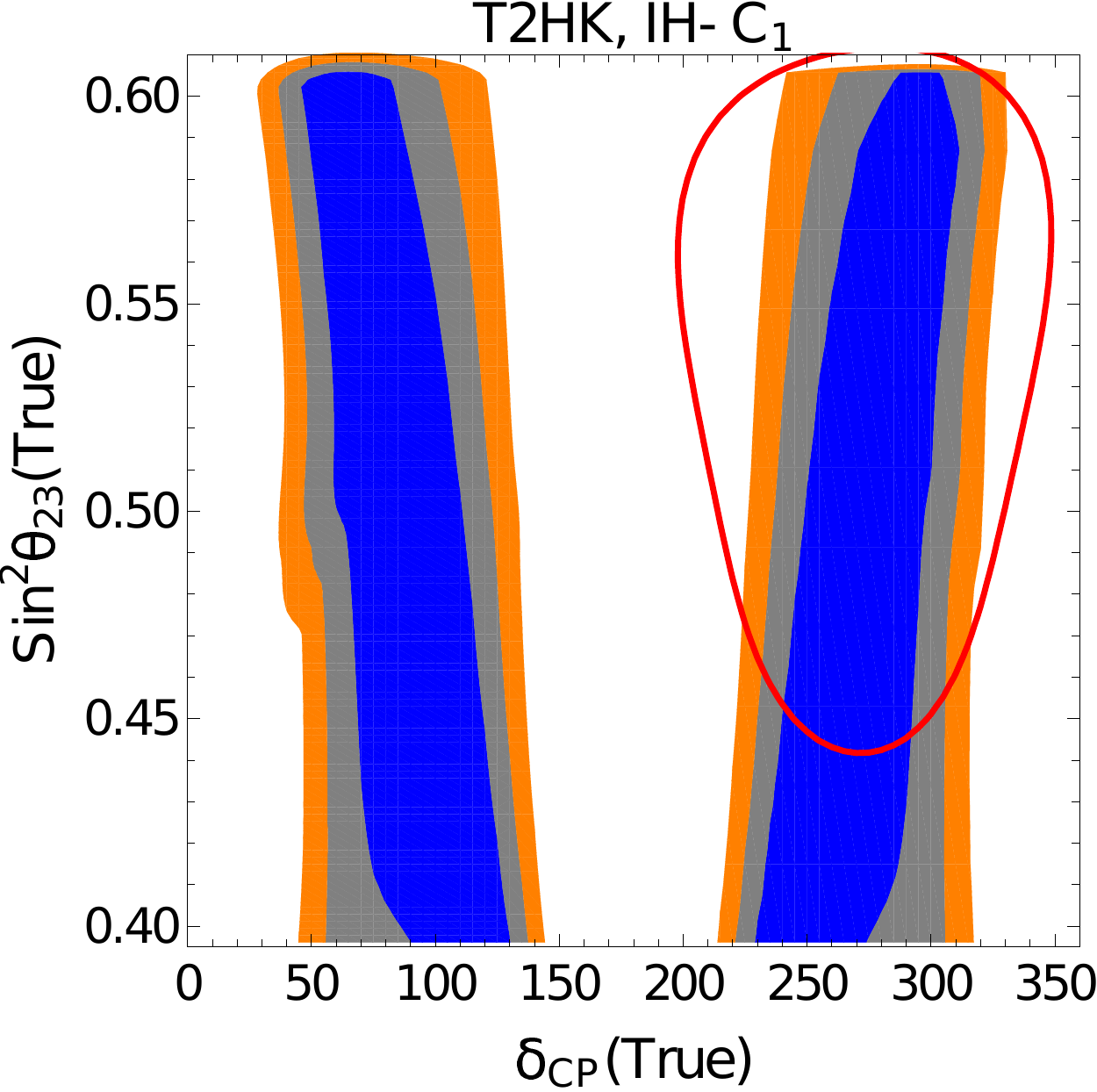}
\hspace{0.5in}
\includegraphics[height=5.8cm,width=6cm]{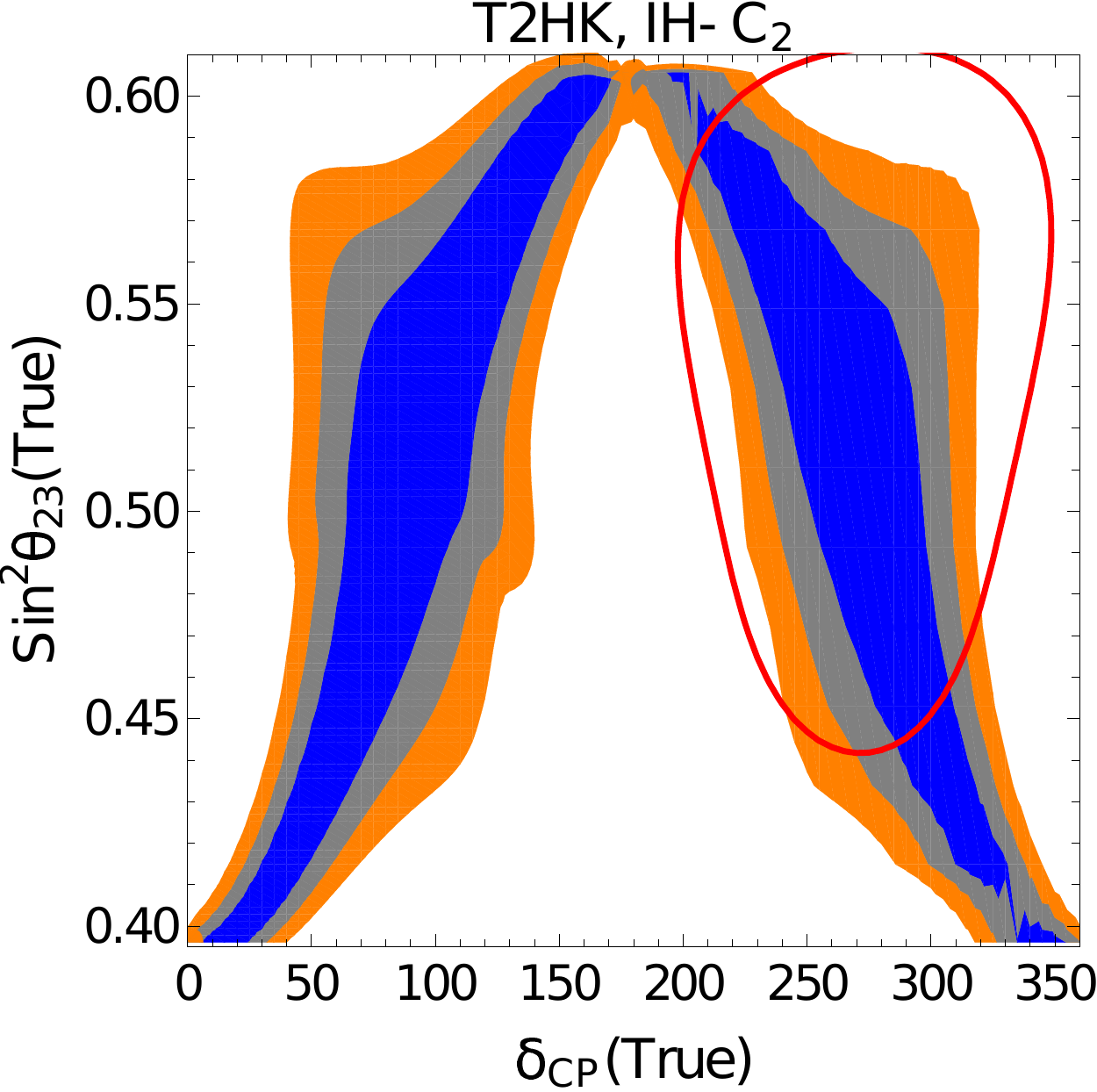} \\
\includegraphics[height=5.8cm,width=6cm]{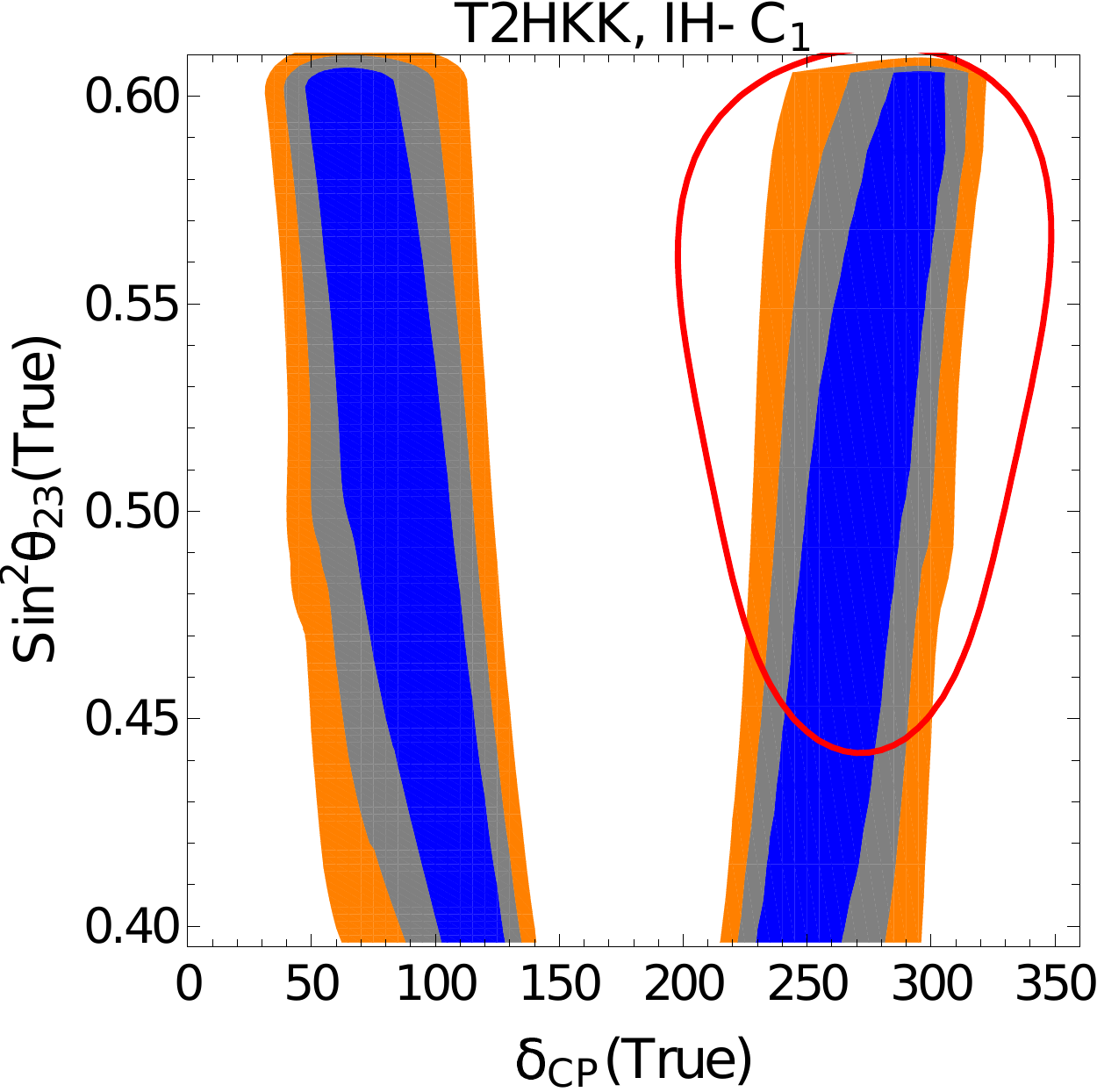}
\hspace{0.5in}
\includegraphics[height=5.8cm,width=6cm]{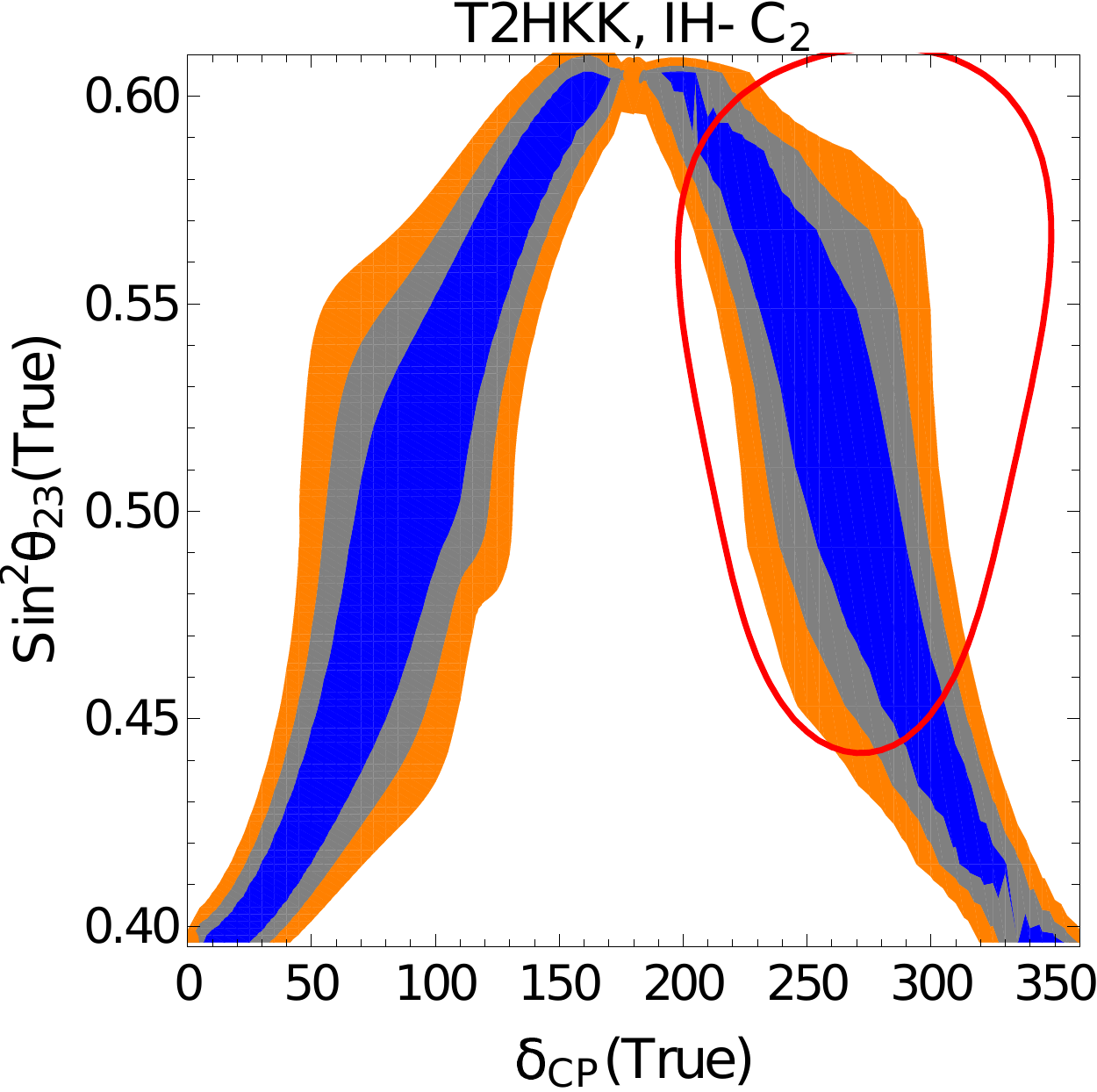} \\
 \end{tabular}
 \end{center}
\vspace{-0.8cm} 
\caption{Same as fig.~(\ref{fig:th23_cp_bound}) but for Inverted Hierarchy.}
\label{fig:th23_cp_bound-IH}       
\end{figure}

The resultant plots are shown in  fig.~(\ref{fig:th23_cp_bound}) for true hierarchy as NH 
and fig.~(\ref{fig:th23_cp_bound-IH}) for true hierarchy as IH.
We assume hierarchy to be known and do not marginalize over hierarchy  
\footnote{We verified that the  effect of marginalizing over the hierarchy is very less. Hence to save  the computation time we have presented the plots by assuming 
 that the hierarchy is known.}.  The blue, grey and the yellow bands
in the figs.~(\ref{fig:th23_cp_bound}, \ref{fig:th23_cp_bound-IH})
represent $1\sigma$, $2\sigma$, $3\sigma$ regions in the 
$\sin^2\theta_{23} - \dcp$ plane, respectively. The red contours show the  $3\sigma$ allowed area
obtained by the Nu-fit collaboration \cite{Esteban:2016qun,nufit16}. 
These plots show the extent to which these three experiments can test the 
correlations between the two yet undetermined variables $\sin^2\theta_{23}$ and $\dcp$ 
in conjunction with the symmetry predictions. The red contours show the  $3\sigma$ allowed area
obtained by the Nu-fit collaboration \cite{Esteban:2016qun,nufit16}. 
The topmost panel corresponds to DUNE - 40 kT detector whereas the middle and
the lowest panels correspond to T2HK and T2HKK experiments respectively.
The left plots in all the rows are for testing $C_1$  
whereas the right plots are for testing $C_2$.
  
The figures show that, because of the  correlations predicted by symmetries,
certain combinations of the true $\theta_{23}$ and $\dcp$ values get 
excluded by DUNE, T2HK and T2HKK. Owing to their high sensitivity to determine CP violation, T2HK and T2HKK constrain the range of $\dcp$ better than that of DUNE. This can be seen from 
the figures (see figs.~(\ref{fig:th23_cp_bound}, \ref{fig:th23_cp_bound-IH}) which show that, 
as we go from top to bottom the contours gets thinner w.r.t. $\dcp$. For instance for the $C_1$ correlation, the CP conserving values $0^\circ$ and 
$360^\circ$ get excluded at $3\sigma$ for both the octants 
by all the three experiments as can be seen from the plots in the left panels.
However, for the $C_2$, these values are allowed at $3\sigma$ 
for all the three experiments. 
Whereas, $\dcp=0^\circ$ and $360^\circ$ are excluded 
by DUNE and HK experiments at $1\sigma$ and $2\sigma$ respectively. 
Again one can see from the right panels that 
for  $C_2$, $\dcp=180^\circ$ is allowed for $\sin^2\theta_{23} > 0.55$ (i.e. higher octant) by DUNE but gets barely excluded at $2\sigma$ by T2HK and T2HKK experiments.
The correlations predicted by the symmetry considerations 
being independent of hierarchy, the allowed regions are not very different for NH and IH. 
But the region of parameter space allowed by current data for IH 
is more constrained and the symmetry predictions  
restrict it further as can be seen from fig.~(\ref{fig:th23_cp_bound-IH}). 
Some of the  parameter space allowed by the current data can also be
disfavored by incorporating the correlations due to symmetry relations.

\begin{figure}[h!]
\begin{center}
\vspace{0.2in}
\includegraphics[height=12cm,width=14cm,angle=0]{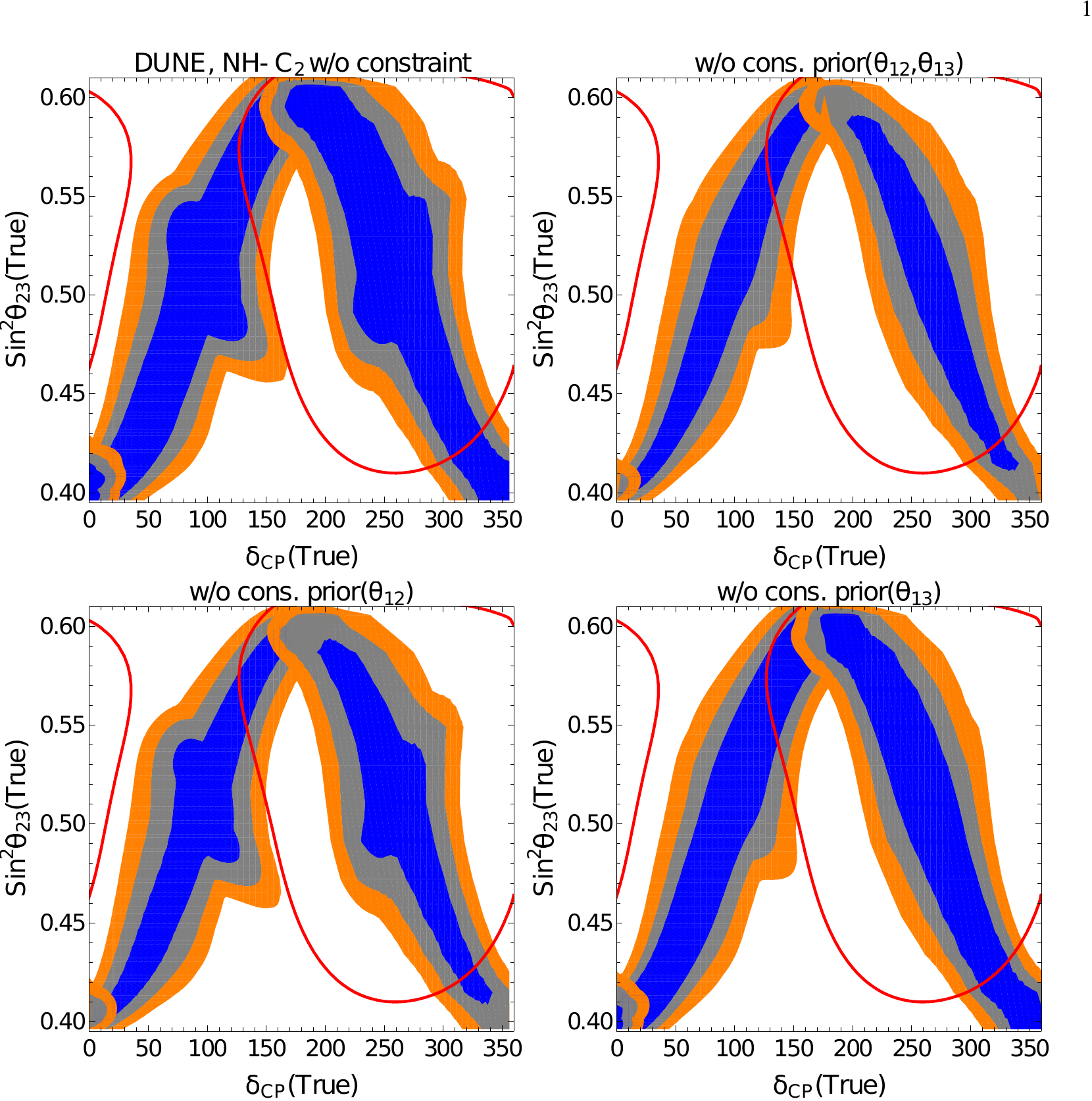} 
\end{center}
\caption{\footnotesize Contour plots in the true: $\sin^{2} \theta_{23}$(true) - $\dcp$(true) plane for DUNE  assuming the symmetry relation $C_2$ in the test. The additional constraints, eq.~(\ref{addn_predictn2}), has not been applied in generating this plot. 
The first  panel represents the plot without including any prior. 
The second panel in the first row shows the effect of prior on $\theta_{12}$ 
and $\theta_{13}$. The first plot in the second row shows the effect of 
inclusion of $\theta_{12}$ prior whereas the second plant shows the 
effect of inclusion of $\theta_{13}$ prior.  
The hierarchy is fixed as NH. The red contour in each panel represents the $3\sigma$ allowed area of the Nu-fit collaboration.  
}
\label{fig:th23_cp_bound-without-const}       
\end{figure}

In section \ref{sec:part_symm}, it was discussed that the additional restrictions, eq.~(\ref{addn_predictn1}) and eq.~(\ref{addn_predictn2}), are obtained when the partial $\mu - \tau$ symmetry is derived in the specific approach discussed. However, possibilities exist where partial $\mu - \tau$ symmetry can be generated without the additional restrictions. 
In this context, we analyzed the changes in the allowed areas when the 
additional restrictions are not imposed. This is done only for DUNE 
which captures the essential trend of the impact of not imposing the 
extra constraints. This is shown in fig.~(\ref{fig:th23_cp_bound-without-const}). 
We have studied this for the representative case of the symmetry relation
$C_2$. The procedure for generating the plots is the same as outlined earlier excepting the 
inclusion of priors. While the earlier plots were generated without any prior,
for these cases, we have studied the following scenarios: 
\begin{enumerate}
\item No prior on $\theta_{13}$ and $\theta_{12}$.
\item Prior on $\theta_{13}$ and $\theta_{12}$.
\item Prior on $\theta_{12}$ and no prior $\theta_{13}$.
\item Prior on $\theta_{13}$ and no prior  $\theta_{12}$.
\end{enumerate}
The first plot of top row is without any prior on the parameters $\theta_{12}$ and 
$\theta_{13}$ and no additional constraints imposed. We find that the allowed 
area increases in size as compared to the cases where the extra 
constraints embodied in eq.~(\ref{addn_predictn1}) and eq.~(\ref{addn_predictn2})
are not imposed. 
The second plot of top row is without imposing these additional restrictions,
but including prior on $\theta_{12}$ and $\theta_{13}$. 
In this case the allowed regions are more restricted and certain combinations
of $\theta_{23}$ and $\dcp$ get disfavoured. We proceed further to show the impact of prior considering a single mixing angle at a time in the second row.
In the first plot of the bottom row, we show the effect of including prior on 
$\theta_{12}$ but no prior on $\theta_{13}$. In this case the shape of 
the allowed regions are same but they reduce in size. The effect of 
certain combination of $\theta_{23}$ and $\dcp$ values getting disfavoured
are seen more at the $1\sigma$ level.  Similarly, the second plot of the bottom row shows the effect of $\theta_{13}$ prior. In this case also, the allowed regions reduce in size 
as compared to the case where no priors are included (first panel of top row). 

\subsection{Differentiating between the  $ C_1 $ and $ C_2$ symmetries }\label{sec:result_model_diff}
\begin{figure}[h!]
\begin{center}
\begin{tabular}{lr}
 \hspace*{-0.2in}
\includegraphics[height=6cm,width=6cm]{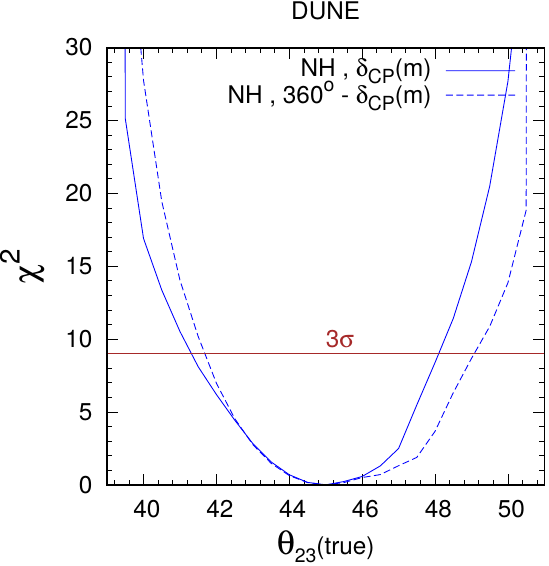} 
\includegraphics[height=6cm,width=6cm]{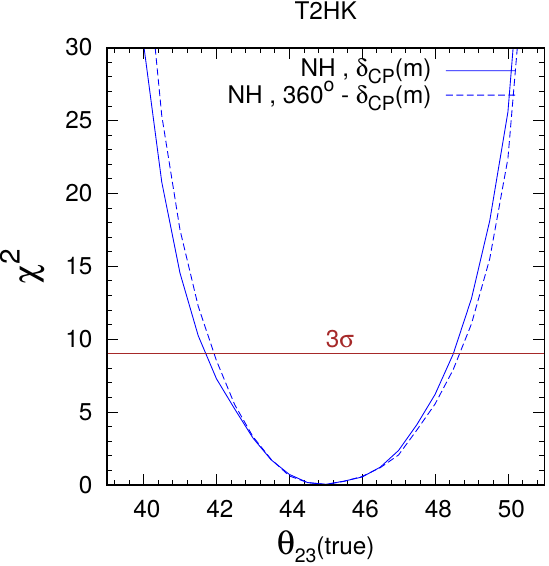}
\includegraphics[height=6cm,width=6cm]{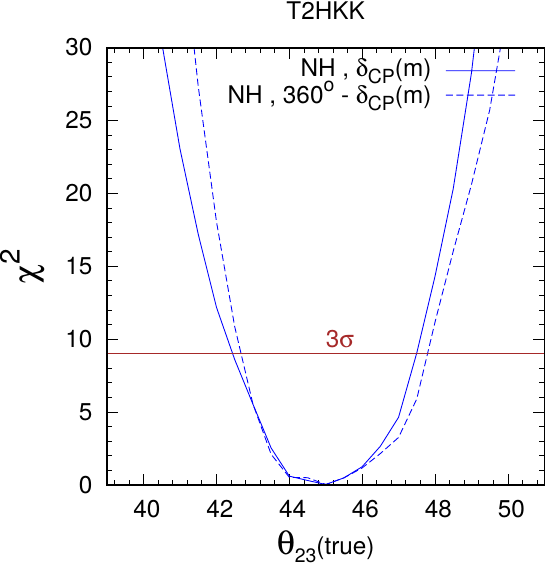} 
 \end{tabular}
 \end{center}
\vspace{-4ex}        
\caption{ The sensitivity of the DUNE, T2HK \& T2HKK experiments to differentiate
between $ C_1 $ and $ C_2$ correlations for known normal hierarchy.} 
\label{fig:diff_z2_z2bar}
\end{figure}

In this subsection, we explore the possibility of differentiation between the symmetries $ C_1 $ and $ C_2$. This is presented in fig.~(\ref{fig:diff_z2_z2bar})
where we plot $ \Delta \chi^{2} $ vs  true $\theta_{23}$.
To find $ \chi^{2}_{stat}$ (as defined in eq.~(\ref{eq:stat_chisq})),
true events are calculated by  varying the true values of $\theta_{23}$ in the range ($ 39^\circ - 51^\circ $). For each true $\theta_{23}$,  true values of $\sin^2\theta_{13}$ and $\sin^2\theta_{12}$ are allowed to vary in their $3\sigma$ range such that the condition  as given in eq.~(\ref{addn_predictn1}) is satisfied. Using these true values of the angles 
the true $ \delta_{CP} $ values are calculated using the correlation $C_1$. 
This leads to two sets of true events corresponding to $\dcp$ and ($360^\circ - \dcp$), respectively. 
The remaining oscillation parameters are kept fixed at their best-fit values as shown in table \ref{tab:param_values}. In the theoretical fit, to calculate test events, we marginalize  over
 $\sin^2\theta_{13}$, $|\Delta m^2_{31}|$,
$\sin^2\theta_{23}$ in the range given in table~\ref{tab:param_values}
and test $ \delta_{CP} $ values are calculated using the $C_2$.  In addition we impose the condition as given in eq.~(\ref{addn_predictn2}) connecting test $\sin^2\theta_{23}$ and $\sin^2\theta_{13}$ and compute the $\chi^2$. For each choice of true $\sin^2\theta_{23}$, the $\chi^2$ 
is marginalized over true $\sin^2\theta_{13}$ and  $\sin^2\theta_{12}$ and 
the minimum $\chi^2$ for each true $\sin^2\theta_{23}$ is taken as 
the value of $\chi^2$. This process is done for both $\dcp$ and $360-\dcp$ separately. 
We  have performed the  analysis considering the true hierarchy as NH.
We have checked that if we assume IH as the true hierarchy we obtain
similar results. 
The three panels from left to right represent DUNE, T2HK and T2HKK respectively. 
The solid blue curves in the plots are for predicted range $\dcp \in (0^\circ < \dcp< 180^\circ)$ and the dashed blue curves in the plots are for complementary range $360^\circ - \dcp \in (180^\circ < \dcp< 360^\circ)$ as predicted by the correlations.
The brown solid line shows the 3$\sigma$ C.L.
We observe from the figure that at  maximal $ \theta_{23} $  both the correlations are indistinguishable by all the three experiments as is expected from the eqs.~(\ref{eq:model_1}, \ref{eq:model_2}). 
%

 \begin{table}[h!]
\centering \scriptsize
  \begin{tabular}{|c|c|c|c|c|c|c|}
 \hline
 \multirow{1}{*}{Range of $\delta_{CP}$} &  \multicolumn{2}{c|}{DUNE} & 
    \multicolumn{2}{c|}{T2HK} & \multicolumn{2}{c|}{T2HKK}\\
    \cline{1-7}
$0^\circ \leq \delta_{CP} \leq 180^\circ$ & $\theta_{23} \leq 41.5^\circ$ &  
$\theta_{23} \geq 48^\circ$ &   $\theta_{23} \leq 41.8^\circ$ & $\theta_{23} \geq 48.5^\circ $ &$ \theta_{23} \leq 42.6^\circ$ & $\theta_{23} \geq 47.5^\circ $ \\
\hline
$180^\circ \leq \delta_{CP} \leq 360^\circ$ & $\theta_{23} \leq 41.8^\circ$ & $\theta_{23} \geq 49^\circ $ & $\theta_{23} \leq 42^\circ$ & $\theta_{23} \geq 48.7^\circ $ & $\theta_{23} \leq 42.8^\circ$ & $ \theta_{23} \geq 47.7^\circ $ \\
\hline
\end{tabular}
 \caption{\footnotesize The  limits 
of $\theta_{23}$ in degrees below and above which the correlations $C_1$ and $C_2$ can be 
differentiated at 3$\sigma$ C.L. for two different ranges of $\delta_{CP}$.
}
 \label{tab:mod-diff}
 \end{table}

The capability of the experiments to differentiate between the two correlations 
increases as we move away from maximal value. 
The range of $\theta_{23}$ for which the three experiments can differentiate between the correlations at $3\sigma$ is given in table \ref{tab:mod-diff}. The lower limits signify the values of $\theta_{23}$ below which the correlations can be differentiated at $3\sigma$ and the upper limits is for the values above which the same can be achieved.

\section{Conclusion}\label{sec:conclusion}
We study here  partial $ \mu - \tau $ reflection symmetry of the leptonic mixing matrix, $U$, which can arise from discrete flavor symmetry. Specific assumptions which lead to this symmetry were reviewed here. This symmetry implies $ |U_{\mu i}|=|U_{\tau i}|~(i=1,2,3) $ for a single column of the leptonic mixing
matrix $U$. If this  is true for the third column of $U$ then it
leads to maximal value of the atmospheric mixing angle and CP phase $ \dcp $.
However, if this is true for the first or the second column 
then one obtains definite correlations among  $\theta_{23} $ and $\delta_{CP}$.
We call these scenarios $C_1$ (equality for the first column) 
and ${C_2}$ (equality of the second column).   
We find that almost all the discrete subgroups of SU(3), except a few exceptional cases, having three dimensional irreducible representations display the form of partial $ \mu - \tau $ symmetry. We study the correlations among  $ \theta_{23} $ and $\delta_{CP}$ in the two scenarios. 
Each scenario gives two values of $\delta_{CP} $ for a given 
$\theta_{23} $ -- one belonging to $0^\circ < \dcp < 180^\circ$  
and the other belonging to  $180^\circ < \dcp < 360^\circ$. 
The models also give specific correlations between $\theta_{23}$ and $\dcp$ 
and these are opposite for $C_1$ and $C_2$. 
We study how the allowed areas in the $\sin^2\theta_{23} - \dcp$ plane 
obtained by the global analysis of neutrino oscillation data from the Nu-Fit collaboration 
compare with the predictions from the symmetries. 

We also expound the testability of these symmetries considering next generation accelerator based  experiments, DUNE and Hyper-Kamiokande.
This is illustrated in terms of plots in the $\sin^2\theta_{23}$ (true) - $\dcp $(true) plane obtained by fitting the simulated experimental data  with 
the symmetry predictions for $\dcp$.   
The values of $\theta_{23}$ are found to be more constrained for 
the CP conserving values namely $\dcp = 0^\circ, 180^\circ, 360^\circ$. 
For the $C_2$ correlation, the $\theta_{23}$ is found to be in the higher octant 
for $\dcp = 180^\circ$ and in the lower octant for $\dcp = 0^\circ$ and $360^\circ$. 
For the correlation $C_1$, values of $\dcp$ around all the three CP conserving values $\dcp =0^\circ$, $180^\circ$ and $360^\circ$ are seen to be disfavored.
Finally, we illustrate the capability of DUNE and Hyper-Kamiokande to distinguish between the predictions of the two correlations. 
We observe that both the experiments can  better differentiate between these two  as 
one moves away from the  maximal $\theta_{23} $ value. 

In conclusion, the future experiments provide testing grounds for 
various symmetry relations, specially those connecting $\theta_{23}$ and 
$\delta_{CP}$. 


\acknowledgments
The research work of ASJ was supported by BRNS (Department of Atomic Energy) and by Department of
Science and Technology,  Government of India through the Raja Ramanna fellowship and
the J. C. Bose grant respectively. NN gratefully acknowledges the support in part by the National
Natural Science Foundation of China under grant No.11775231 for the research work. 


\bibliography{mutau-flav-ref.bib}

\end{document}